\documentclass[lettersize,journal]{IEEEtran}
\usepackage{amsmath,amsfonts}
\usepackage{algorithmic}
\usepackage{algorithm}
\usepackage{soul}
\usepackage{array}
\usepackage[caption=false,font=normalsize,labelfont=sf,textfont=sf]{subfig}
\usepackage{textcomp}
\usepackage{stfloats}
\usepackage{float}
\usepackage{url}
\usepackage{verbatim}
\usepackage{graphicx}
\usepackage{cite}
\usepackage{amsmath}
\usepackage{tikz}
\usepackage{xcolor}
\usepackage{pgfplots}
\usepackage{standalone}
\usepackage{xcolor}
\usepackage{hyperref}
\definecolor{mycolor1}{rgb}{0.90471,0.19176,0.19882}%
\definecolor{mycolor2}{rgb}{0.29412,0.54471,0.74941}%
\definecolor{mycolor3}{rgb}{0.37176,0.71765,0.36118}%
\definecolor{mycolor4}{rgb}{1.00000,0.54824,0.10000}%
\definecolor{mycolor5}{rgb}{0.86500,0.81100,0.43300}%
\definecolor{mycolor6}{rgb}{0.68588,0.40353,0.24118}%
\definecolor{mycolor7}{rgb}{0.97176,0.55529,0.77412}%
\definecolor{mycolor8}{rgb}{0.63647,0.37529,0.67529}%
\usetikzlibrary{shapes.geometric}
\usetikzlibrary{shadows,shadows.blur}
\usetikzlibrary{arrows.meta}
\pgfplotsset{compat=1.5}

\DeclareMathOperator{\Real}{Re}
\DeclareMathOperator{\opt}{opt}

\DeclareMathOperator{\rank}{rank}
\DeclareMathOperator{\SNR}{SNR}
\DeclareMathOperator{\feasibilityrate}{feasibility \ rate}
\DeclareMathOperator{\bpsph}{bits/sec/Hz}
\DeclareMathOperator{\bpsphpu}{bits/sec/Hz/user}
\DeclareMathOperator{\dB}{dB}
\DeclareMathOperator{\Tr}{Tr}
\DeclareMathOperator{\Span}{span}
\DeclareMathOperator{\vect}{vec}
\DeclareMathOperator{\SINR}{SINR}
\DeclareMathOperator{\ISMR}{ISMR}
\DeclareMathOperator*{\argmax}{arg\,max}

\newcommand{\eqdef}{\overset{\Delta}{=} }

\newcommand*\MyScale{1}

\tikzset{every picture/.style={scale=\MyScale}}

\begin{document}

\title{On Outage-based Beamforming Design for Dual-Functional Radar-Communication 6G Systems}

\author{Ahmad Bazzi and Marwa Chafii
\thanks{Ahmad Bazzi is with the Engineering Division, New York University (NYU) Abu Dhabi, 129188, UAE
(email: \href{ahmad.bazzi@nyu.edu}{ahmad.bazzi@nyu.edu}).\\ 

Marwa Chafii is with Engineering Division, New York University (NYU) Abu Dhabi, 129188, UAE and NYU WIRELESS,
NYU Tandon School of Engineering, Brooklyn, 11201, NY, USA (email: \href{marwa.chafii@nyu.edu}{marwa.chafii@nyu.edu}).}
\thanks{}}

\markboth{\MakeLowercase{to appear in} IEEE Transactions on Wireless Communications, 2023}%
{Shell \MakeLowercase{\textit{et al.}}: A Sample Article Using IEEEtran.cls for IEEE Journals}


\maketitle

\begin{abstract}
This article studies and derives beamforming design in a dual-functional radar-communication (DFRC) multiple-input-multiple-output system. We focus on a scenario, where the DFRC base station communicates with downlink communication users, with imperfect channel state information knowledge, and performs target detection, all via the same transmit signal. Through careful relaxation procedures, we arrive at a suitable and novel optimization problem, which maximizes the radar output power in the Bartlett sense, under probabilistic outage signal-to-interference-and-noise ratio constraints. Theoretical analysis proves optimality of the solution given by the relaxed version of the problem, as well as closed-form solutions in certain scenarios. Finally, the achieved performances and trade-offs of the proposed beamforming design are demonstrated through numerical simulations.
\end{abstract}

\section{Introduction}
\IEEEPARstart{6}{G} is expected to nurture a wide scope of services, spanning haptic telemedicine to VR/AR remote services and holographic teleportation with massive eXtended reality (XR) capabilities, as well as Blockchain \cite{blockchain}, just to name a few. The benefits of $6$G go beyond communication. It will incorporate new capabilities like sensing and computing, allowing new services and utilizing improved environmental information for machine learning and artificial intelligence. On the other hand, such bandwidth-hungry applications require a $10^3 \times$ capacity increase, as highlighted by \cite{saad}. Even more, the monthly mobile data per mobile broadband subscriber is expected to grow from $5.3$ GB, by $2020$, to $257$ GB by $2030$, which is a $50\times$ increase within a decade \cite{6g-the-new-horizon}.

Indeed, integrated sensing and communications cooperation \cite{ISAC-cooperation} and convergence is an utmost topic that will serve as a door-opener for innovative applications like never seen before, such as autonomous driving \cite{autonomous-vehicules}, robotics and UAV sector. Consequently, coexistence  between radar and communication systems \cite{ISAC-coexistance} has been a primary field of investigation\footnote{Not to be mistaken with radar-communication coexistence, which is distinct from integrated sensing and communication. See \cite{ISAC-seperate} for more details.}. Joint communication and sensing comes with a number of practical advantages, an obvious one being spectrum efficiency, due to the interoperability between the radar and communication counterparts. Another advantage on the table is hardware resource sharing between radar and communication tasks, thus having both on a single platform will lead to reduced PHY-layer modem size and cost. With these merits come a serious number of challenges and research questions that need to be addressed, such as efficient resource reuse and spectrum sharing \cite{isac-spectrum-sharing-02}, trade-offs between high communication rates and high-resolution sensing performances \cite{isac-tradeoffs-01,isac-tradeoffs-02,isac-tradeoffs-03}, privacy and security \cite{Secure}, shared waveform design \cite{ISAC-waveform-design-01,ISAC-waveform-design-02,toward-dfrc}. In this article, we focus on beamforming design, so that the beamformed signal is optimized for both communication and sensing tasks.

Heretofore, previous beamforming design for joint radar and communications have been conducted under perfect channel state information (CSI). Even though CSI may be estimated between the dual-functional radar-communication (DFRC) base station and the communication users\footnote{CSI acquisition may be done via one-way where users send a training sequence assuming channel reciprocity, or two-way where DFRC base station sends a sequence, then channel estimation is done per user and fed back to the base station}, the CSI cannot be perfect. A main driver of this inaccurate knowledge comes from hardware imperfections \cite{HW-imperfect}. For instance, quantization errors which are caused by analog-to-digital converters (ADCs) within the RF-Frontend of the PHY modem, will further lead to estimation errors, which in turn leads to fixed-point error propagation. From a hardware perspective, it is highly desired to economize the total bit width of the baseband design. This need serves as key motivation behind sophisticated baseband modems, that are aware of imperfect CSI. Joint communication and sensing DFRC modems will not be an exception.

\subsection{Existing work}
Robust beamforming is a well-known and interesting topic for generic MIMO communication systems, when partial CSI is available. The work in \cite{cognitive-radio} formulates a robust beamforming design aiming at an optimization problem for interference control in a cognitive radio setting, i.e. a communication optimized scenario with no radar functionality. Furthermore, another communication-only robust beamformer is derived in \cite{Secrecy1} in the context of security of MIMO wiretap systems. The goal is to minimize the $\SINR$ towards an eavesdropper, while ensuring a target $\SINR$ for the users, in the presence of imperfect CSI. The problem in \cite{Secrecy2} is dedicated to downlink systems, where the goal is to minimize the transmit powers while outage probabilities with imperfect CSI towards eavesdropper. Meanwhile, an intelligent reflecting surface aided robust beamforming design is proposed in \cite{robust-IRS} for communications-only, to optimize for transmit power, under per-user quality of service constraints. Another class of robust beamforming assume a deterministic approach \cite{conservative}, where the unknown CSI is assumed to be bounded by a deterministic quantity, hence worst-case analysis \cite{palomar,worstcase2}. Deterministic robust beamforming has been applied to contexts, such as high-mobility downlink scenarios \cite{Noma-outage}. In \cite{GaussianRandomization}, some methods were studied in the context of robust beamforming, for example sphere bounding and Bernstien-type inequality based beamforming. In all methods within \cite{GaussianRandomization}, the main objective is to minimize the total transmit beamforming power under some QoS constraint per communication user. We highlight the following main differences: (i) Our method is concerned with maximizing the total power at the output of the Bartlett beamformer for radar purposes, rather than minimizing the total transmitted beamforming power (ii) Our method bounds the total transmit beamforming power by the available power budget, instead of minimizing it. 
\textit{Unlike previous work on generic MIMO robust beamforming, whether stochastic or deterministic robust design,  we incorporate radar functionality as an intrinsic part of the robust design of the DFRC framework, enabling the DFRC base station to illuminate beams towards intended targets, while communicating with users with imperfect CSI.}
In the context of DFRC, the work in \cite{toward-dfrc} studies waveform design for DFRC systems, where the proposed method is a two-step approach: In the first step, one fabricates a radar-only waveform given a desired beampattern to generate a radar-only waveform. In the second step, one solves a weighted combination optimization problem designed to trade-off communication and radar costs, given a power budget on the transmitted waveform. Moreover, \cite{toward-dfrc} does not consider an imperfect CSI-based approach. In contrast to \cite{toward-dfrc}, our method is a single-step approach assuming imperfect CSI knowledge. Besides \cite{toward-dfrc} does not assume clutter within the radar sub-model. Furthermore, the work in \cite{LimitedFeedforward} formulates a power allocation problem over OFDM subcarriers by first minimizing the Cramer-Rao bound on delay-doppler estimation to achieve a radar-only waveform. The next step entails optimizing the power allocated per OFDM subcarrier through minimizing the error on the linear minimum mean square estimate (LMMSE) and a similarity constraint on the radar-only waveform. The imperfect CSI comes from imperfect knowledge of transmit powers, only. In this article, we focus on beamforming rather than power allocation over OFDM subcarriers, in the presence of channel uncertainty rather than transmit power uncertainty. Furthermore, our approach consists of a single optimization framework, as opposed to that in \cite{LimitedFeedforward}. In addition, we assume that delay-doppler estimation for radar waveforms is out of the scope of this paper, as the DFRC implements an estimation block after having received the radar echo. The method described in \cite{Secure} is a security-related DFRC beamforming problem focusing on minimizing the signal-to-noise ratio towards an eavesdropper, with some partial knowledge on the position of the eavesdropper. 
\textit{Unlike previous works on joint communications and sensing that consider perfect CSI with deterministic $\SINR$, we propose in this paper a new optimization framework for DFRC taking into account outage $\SINR$ probability. To the best of our knowledge, outage $\SINR$ probability considerations, and therefore the achievable trade-offs, in the context of joint sensing and communications have not been addressed, yet.} 
In the following two subsections, we highlight novel contributions and shed light on some important insights appearing in this manuscript.

\subsection{Contributions and Insights}
The paper considers dual-function radar communication beamforming towards communication users, while scanning targets in specific directions. Due to the imperfect CSI case, our precoding scheme aims at optimizing weight vectors to guarantee a maximal outage $\SINR$ level, while maximizing the power of the radar beamformer in the desired look-direction, through the Bartlett criterion. The initial problem involves stochastic constraints, which are then transformed to deterministic forms through a series of carefully chosen relaxations. To this end, we summarize our contributions as follows

\begin{itemize}

        \item \textbf{DFRC beamforming design under imperfect CSI}. Under practical scenarios where CSI is not fully known to the DFRC base station, we propose a beamforming design framework tailored for integrated sensing and communication systems with partial access to CSI knowledge. Furthermore, a suitable optimization problem is proposed. However, due to the intractability of the optimization problem at hand, we leverage tools from convex optimization theory, such as \textit{Bernstien-type} inequalities, in order to bypass probabilistic event constraints, \textit{Schur complement} properties onto auxiliary variables, and rank relaxation techniques so as to arrive at a tractable convex optimization problem. To this end, we make use of classical optimization solvers to solve the final DFRC beamforming.
	\item \textbf{Closed-form solutions of DFRC beamforming design}. The derived closed-form beamforming solutions for the single-user (SU) case are derived and analyzed. Thanks to these expressions, we provide novel insights into trade-offs between the communication and radar subsystems under imperfect CSI. More specifically, we highlight the importance of each parameter involved in the closed-form expressions, and their role in tuning a desired beamforming solution.

    \item \textbf{Extensive simulation results}. Finally, we perform extensive numerical simulations to corroborate the robustness of the DFRC beamforming design, by illuminating both communication and sensing capabilities. Moreover, we illustrate the superiority of the proposed DFRC beamforming, against the classical sphere bounding and Bernstein designs described in \cite{GaussianRandomization} in terms of minimum achievable rate and under imperfect CSI scenarios. We also show how the achievable communication rates affect radar sensing metrics, under imperfect CSI.
	
\end{itemize} 
Furthermore, we reveal some important insights, i.e.
\begin{itemize}
    \item  The final relaxed version of the beamforming optimization problem is proved to give a $\rank-$one beamforming solution, therefore avoiding the integration of post-processing procedures, such as Gaussian randomization and principle component analysis \cite{PCA}.
	\item Closed-form solutions reveal the explicit dependencies of outage parameters on the radar-communications trade-off. The level of correlation between the steering vector towards the target and the channel between the DFRC base station and the communication user also has a direct impact on this trade-off. Graphical interpretations are also given to illustrate this point.
	\item The accuracy of the resulting radar's beampattern in the desired look-direction depends on the $\SINR$ outage threshold, in the presence of imperfect CSI. In other words, the higher the $\SINR$ we target, the less capable the radar is in looking at certain directions. Likewise, the radar's sidelobe rejection also depends on the target $\SINR$.
	\item Under a unit transmit power constraint, the proposed design could achieve an average sum-rate of $4.78 \bpsphpu$ with a probability of detection of $\sim 0.99$, as compared to $\sim 0.65$ when state-of-the-art methods are adopted \cite{toward-dfrc}. Interpretation on such radar gains are given in the simulations section.
	\item For a fixed $\ISMR$ level, the average achievable rate increases with a decrease in number of communication users or an increase in the desired $\SINR$ outage threshold.
\end{itemize}
\subsection{Organization and Notations}
The rest of this paper is organized as follows. We introduce the system model for radar and communications, as well as some KPIs, in Section \ref{sec:system-model}. The proposed DFRC beamforming design with imperfect CSI is derived in Section \ref{sec:beaforming-design}. Section \ref{sec:simulations} presents extensive simulation results, and Section \ref{sec:conclusion} concludes the paper. \\\\
\textbf{Notation}: Upper-case and lower-case boldface letters denote matrices and vectors, respectively. $(.)^T$, $(.)^*$ and $(.)^H$ represent the transpose, the conjugate and the transpose-conjugate operators. The statistical expectation is $\mathbb{E}\lbrace . \rbrace$. The vectorization and unvectorization operators are denoted as $\vect$ and $\vect^{-1}$, respectively. In particular, $\vect$ takes an $N\times M$ matrix $\pmb{X}$ as input and returns an $NM \times 1$ vector, by stacking the columns of $\pmb{X}$. For any complex number $z \in \mathbb{C}$, the magnitude is $\vert z \vert$, its angle is $ \angle z$, and its real part is $\Real(z)$. The Frobenius norm of matrix $\pmb{X}$ is $\Vert \pmb{X} \Vert$. We denote a positive semi-definite matrix as $\pmb{X} \succeq \pmb{0}$ and $\lambda_{\max}(\pmb{X})$ is the maximum eigenvalue of matrix $\pmb{X}$. The matrix $\pmb{I}_N$ is the identity matrix of size $N \times N$. The zero-vector is $\pmb{0}$. The $\rank$ returns the rank of a matrix and $\Tr$ returns the trace of a matrix. The inverse of a square matrix is $\pmb{X}^{-1}$. The probability of an event $\mathcal{A}$ is $\Pr(\mathcal{A})$. The span of vectors $\pmb{x}_1 \ldots \pmb{x}_N$ is $\Span \lbrace \pmb{x}_1 \ldots \pmb{x}_N \rbrace$. The noncentral chi-squared distribution function with $n$ degrees of freedom and parameter $\rho$ is denoted as  $\mathcal{F}_{\chi^2_n(\rho)}$ and the central chi-squared distribution with $n$ degrees of freedom is $\mathcal{F}_{\chi^2_n}$.

\section{System Model}
\label{sec:system-model}
	Let us consider a dual-functional radar communication (DFRC) system comprised of a target of interest, $K$ single-antenna communication users, and a DFRC base station. The base station is equipped with an antenna array composed of $N$ elements and the array response of a signal arriving at direction $\theta$ is denoted as $\pmb{a}(\theta) \in \mathbb{C}^{N \times 1}$, where its $n^{th}$ entry, $a_n(\theta)$ is the phase shift at the $n^{th}$ antenna due to $\theta$. For example, in a uniform linear array setting, we have that 
	\begin{equation}
	    \pmb{a}(\theta) 
	    =
	    \begin{bmatrix}
	        1  & e^{j \frac{2\pi}{\lambda} d \sin(\theta)} & \ldots & e^{j \frac{2\pi}{\lambda} (N-1) d \sin(\theta)}
	    \end{bmatrix}^T,
	\end{equation}
	where $\lambda$ is the wavelength of the signal and $d$ is the inter-element spacing. 

\begin{figure}[!t]
\centering
\includegraphics[width=3.5in]{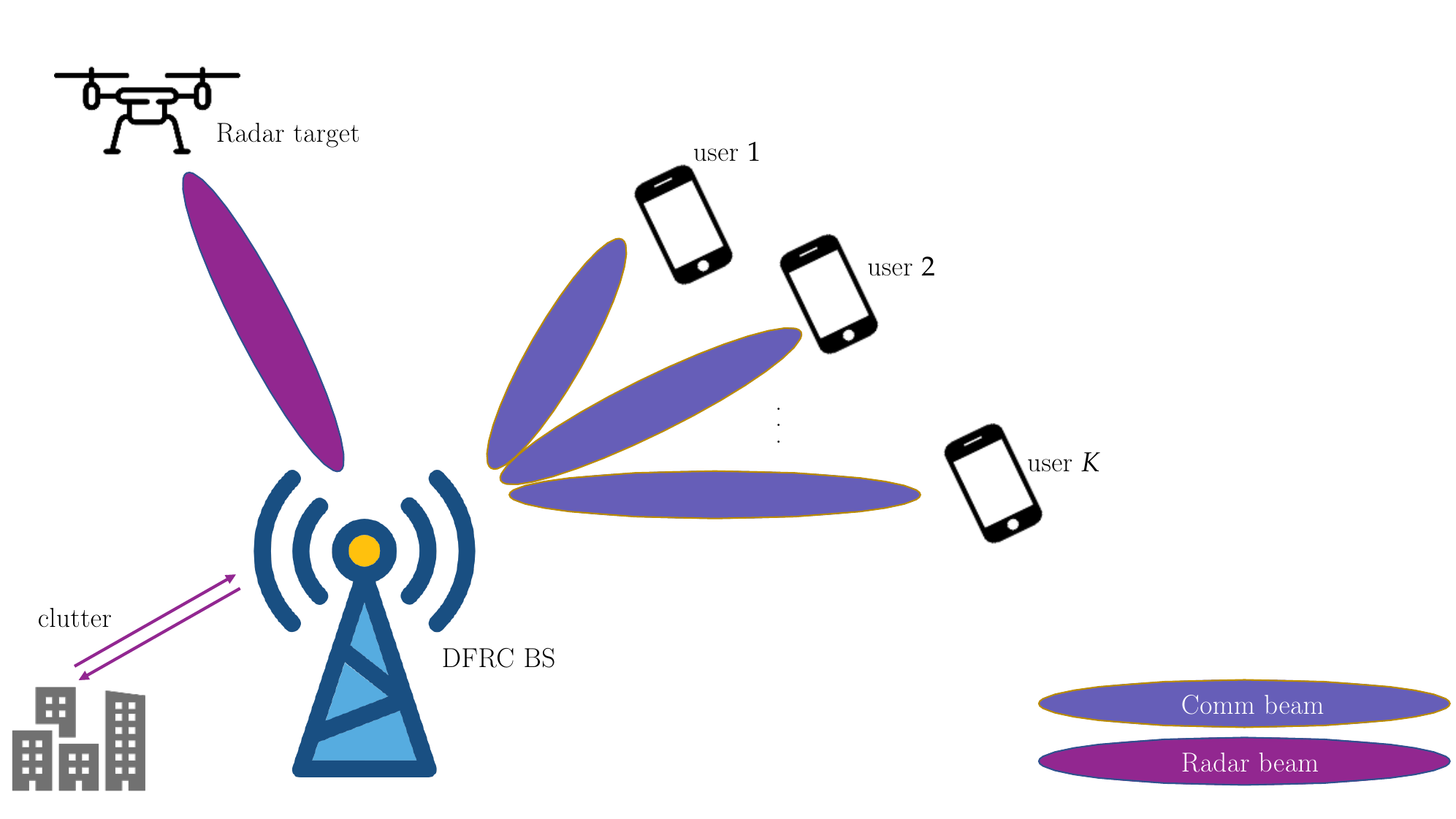}
\caption{
DFRC scenario including a DFRC base station, an intended target with clutter and $K$ communication users.}
\label{fig_1}
\end{figure}
In Fig. \ref{fig_1}, we highlight a use case, where the DFRC base station uses the same signal vector and broadcasts it towards communication users and an intended target of interest. The communication users are assumed to be randomly distributed, whereas the radar beam scan is at a known angle $\theta_0$ from the DFRC base station. Now, let us introduce the communication and radar system models.

\subsection{Communication System Model}
At the $n^{th}$ snapshot, the single transmission in the downlink could be expressed as
\begin{equation}
	\label{eq:y=Hxpz}
	\pmb{y}_c[n] = \widetilde{\pmb{H}}\pmb{x}[n] + \pmb{z}_c[n] ,
\end{equation}
where $\pmb{y}_c[n] \in \mathbb{C}^{K \times 1}$ is the vector of received signals over all communication users, i.e. the $k^{th}$ entry of where $\pmb{y}_c[n]$ is the signal dedicated towards the $k^{th}$ communication user, at the $n^{th}$ snapshot. The channel matrix is given by $\widetilde{\pmb{H}} = \begin{bmatrix} \tilde{\pmb{h}}_1 & \tilde{\pmb{h}}_2 & \ldots & \tilde{\pmb{h}}_K \end{bmatrix}^T \in \mathbb{C}^{K \times N}$, and is flat Rayleigh type fading, assumed to be constant during one transmission. Furthermore, the transmit signal vector is denoted by $\pmb{x}[n]$. Finally, the vector $\pmb{z}_c[n] \in \mathbb{C}^{K \times 1}$ is background noise, assumed to be white Gaussian i.i.d with zero mean and a multiple of identity covariance matrix as $\pmb{z}_c[n] \sim \mathcal{N}(0,\sigma_c^2 \pmb{I}_K)$. Note that the signal $\pmb{x}[n]$ is used for both communication and sensing tasks, hence the vector $\pmb{x}[n]$ is also a snapshot of a radar pulse \cite{toward-dfrc}. 
We further assume that $\pmb{x}[n]$ is precoded, which is the output of the following linear equation
\begin{equation}
	\label{eq:x=Ws}
	\pmb{x}[n] = \pmb{W}\pmb{s}[n] ,
\end{equation}
where $\pmb{W} = \begin{bmatrix} \pmb{w}_1 & \pmb{w}_2 & \ldots & \pmb{w}_K \end{bmatrix} \in \mathbb{C}^{N \times K}$ is the beamforming matrix, where its $k^{th}$ column is the beamforming vector towards the $k^{th}$ user. The vector $\pmb{s}[n] \in \mathbb{C}^{K \times 1}$ contains the QAM symbols intended for all users. In what follows, the symbols are assumed to be independent with unit variance, i.e. $\mathbb{E}(\pmb{s}[n]\pmb{s}^H[n]) = \pmb{I}_K$. Following the concept of ISAC technology, the same signal $\pmb{x}[n]$ will be used for communication, as well as radar sensing. Furthermore, the performance attained by the radar system in our robust beamforming design depends on the independence assumption between the data symbols per user. Note that the independence assumption is a valid one on the data payload portions of transmitted packets in wireless systems, as information bits dedicated per user are independent.

\subsection{Radar System Model}
The radar functionality in the DFRC system should leverage the same transmit signal as the one used for the communication system model, namely $\pmb{x}[n]$. In that way, and using only one time slot, the DFRC is capable of achieving dual sensing and communication functionalities. Thanks to the receiving capability of the DFRC base station, and assuming a colocated mono-static MIMO radar setting, the $n^{th}$ received snapshot, within a certain
time slot, is given as \cite{start-with-this}
\begin{equation}
	\label{eq:yr}
	\pmb{y}_r[n] = 
	\gamma_0 \pmb{a}(\theta_0)\pmb{a}^T(\theta_0)\pmb{x}[n] +
	\sum\limits_{i=1}^I
	\gamma_i \pmb{a}(\theta_i)\pmb{a}^T(\theta_i)\pmb{x}[n] +
	\pmb{z}_r[n].
\end{equation}

We assume that the above reception, sampling and signal processing occur during a time interval termed the coherent processing interval (CPI) \cite{ISAC-waveform-design-02}, which is an interval where sensing parameters (in our case $\theta_0$) remain unchanged. The received signal $\pmb{y}_r[n] \in \mathbb{C}^{N \times 1}$ is composed of the main reflection of the intended target at angle-of-arrival $\theta_0$ with a reflection coefficient $\gamma_0$. The model also accounts for clutter, where the $i^{th}$ clutter is located at $\theta_i$ with reflection coefficient $\gamma_i$. The noise of the radar sub-system is i.i.d modeled as zero mean and a covariance matrix of $\pmb{\Sigma}_{n,r}$. In this work, the main focus is to optimize $\pmb{W}$, according to key performance indicators introduced in the following section. Moreover, all the analysis focuses on a single shared carrier all communication users, as well as the radar.

\subsection{Key Performance Indicators}
To formulate a suitable and robust optimization problem that aims at solving joint sensing and communication problems, it is crucial to define KPIs related to the problem at hand. Based on equations \eqref{eq:y=Hxpz} and \eqref{eq:x=Ws}, the $\SINR$ of the $k^{th}$ user, under the assumption of independent symbols, could be expressed as
\begin{equation}
	\SINR_k
	= 
	\frac{\mathbb{E} \big[\vert \tilde{\pmb{h}}_k^T \pmb{w}_k s_k[n] \vert^2 \big]} 
	     {\sum\limits_{\ell =1, \ell \neq k }^K \mathbb{E} \big[ \vert \tilde{\pmb{h}}_k^T \pmb{w}_{\ell} s_{\ell}[n] \vert^2 \big] + \sigma_{c}^2}.
\end{equation}

In this case, and according to Shannon-theory, the achievable rate of transmission of the $k^{th}$ communication user could be expressed as 
\begin{equation}
\label{eq:rate-equation}
	R_k = \log_2( 1 + \SINR_k ).
\end{equation}
Furthermore, we can write the $\SINR$ of the $k^{th}$ communication user as follows
\begin{equation}
	\SINR_k
	=
	\frac{ \tilde{\pmb{h}}_k ^T \pmb{W}_k \tilde{\pmb{h}}_k^*  } 
	     {\sum\limits_{\ell =1, \ell \neq k }^K \tilde{\pmb{h}}_k^T \pmb{W}_{\ell} \tilde{\pmb{h}}_k^* + \sigma_{c}^2},	
\end{equation}
    where we have assumed unit symbol power, i.e. $\mathbb{E}(\vert s_k \vert^2) = 1$ and $\pmb{W}_k = \pmb{w}_k \pmb{w}_k^H$. Following \cite{GaussianRandomization}, the real channel $\tilde{\pmb{h}}_k$ for each user could be written as $\tilde{\pmb{h}}_k = \pmb{h}_k + \Delta \pmb{h}_k$, where $\pmb{h}_k$ is the estimated channel at the DFRC base station and $\Delta \pmb{h}_k$ is the channel vector error, assumed to be circularly symmetric complex Gaussian as $\Delta \pmb{h}_k \sim \mathcal{N}(\pmb{0},\pmb{C}_k)$, where $\pmb{C}_k$ is the known error covariance matrix. For sake of simplicity, we will assume that $\pmb{C}_k = \sigma_{\Delta h_k}^2 \pmb{I}_N$. This situation is also referred to as imperfect CSI information. Therefore, due to the random nature of the $\SINR$, it makes sense to introduce an outage probability constraint as a QoS indicator of the communication users as 
\begin{equation}
	\Big\lbrace 
	\Pr( \SINR_k \leq \gamma_k ) \leq p_k
	\Big\rbrace_{k=1}^K,
\end{equation}
where $p_k$ is the well-known outage probability used to ensure a desired $\SINR$ level on communication users. Said differently, in average, we can guarantee a minimum $\SINR$ level that is $\gamma_k$ for the $k^{th}$ communication user with a probability of failure equal to $p_k$.\\
On the other hand, the radar should maximize the power in the look-direction of the target, i.e. $\theta_0$. Notice that equation \eqref{eq:yr} directly implies that the covariance matrix of the radar received echo is given as 

\begin{equation}
	\begin{split}
	\mathbb{E}(\pmb{y}_r[n]\pmb{y}_r^H[n])
	&=
	\vert \gamma_0 \vert^2 
	\pmb{a}(\theta_0)
	\pmb{a}^T(\theta_0)
	\mathbb{E}
	\big[
	\pmb{x}[n]
	\pmb{x}[n]^H
	\big]
	\pmb{a}^*(\theta_0)
	\pmb{a}^H(\theta_0)
	 \\
	&+
	2 
	\Real
	\Big[
	\gamma_0
	\pmb{a}(\theta_0)
	\pmb{a}^T(\theta_0)
	\mathbb{E}
	\big[
	\pmb{x}[n]
	\pmb{z}^H[n]
	\big]
	\Big]
	 \\&+
	\mathbb{E}
	\big[
	\pmb{z}[n]
	\pmb{z}^H[n]
	\big],	
	\end{split}
\end{equation}
where $\pmb{z}[n] = \sum\limits_{i=1}^I \gamma_i \pmb{a}(\theta_i)\pmb{a}^T(\theta_i) \pmb{x}[n] + \pmb{z}_r[n]$ is the clutter and noise parts. Focusing on the term of interest, the power due to the signal of interest could be expressed as the trace of the first term as 
\begin{equation}
	\begin{split}
	P(\theta_0)
	&=
	\Tr
	\Big(
	\vert \gamma_0 \vert^2 
	\pmb{a}(\theta_0)
	\pmb{a}^T(\theta_0)
	\pmb{W}\pmb{W}^H
	\pmb{a}^*(\theta_0)
	\pmb{a}^H(\theta_0)
	\Big) \\
	&  \propto
	\sum\limits_{k=1}^K
	\pmb{a}^T(\theta_0)
	\pmb{W}_k
	\pmb{a}^*(\theta_0),
	\end{split}
\end{equation}
where we have used $\mathbb{E}
	\big[
	\pmb{x}[n]
	\pmb{x}^H[n]
	\big] = \pmb{W}\pmb{W}^H$ and $\Vert \pmb{a}(\theta) \Vert^2$ is a constant. The radar sub-system would then aim at maximizing the power in the look-direction, i.e. $P(\theta_0)$. In array processing terms, $P(\theta)$ is the output of a conventional Bartlett beamformer of a signal arriving at angle $\theta$, with covariance $\pmb{W}\pmb{W}^H$. Note that an increase in $P(\theta_0)$ contributes in an increase of the radar's probability of detection, whose expression is given as  \cite{Pd}
\begin{equation}
    \label{eq:probability-of-detection}
    P_D = 1 - \mathcal{F}_{\chi_2^2(\rho)}
    \Big(\mathcal{F}_{\chi_2^2}^{-1}(1 - P_{\text{FA}}) \Big)
\end{equation}
where $\rho = \SNR_r \vert P(\theta_0) \vert^2$, $P_{\text{FA}}$ is the radar's probability of false alarm and $\SNR_r$ is the radar's signal-to-noise ratio. The above KPI is of interest in beamscan applications, where the DFRC base station aims at illuminating a beam in the direction of $\theta_0$ with the intention of maximizing the probability of detection, given a target in that direction $\theta_0$.

\section{Beamforming Design under Imperfect CSI}
\label{sec:beaforming-design}
In a joint sensing and communication approach, the goal is to ideally maximize both radar and communication performances by using the same signal $\pmb{x}[n]$. Alternatively, we can cast the problem as a beamforming design one, where the goal of the DFRC base station is to find the best beamforming matrix that would jointly optimize communication and radar performances. 

\subsection{Problem formulation}
We start by formulating a suitable joint sensing and communication problem, where the main cost function aims at maximizing the output of a Bartlett beamformer subject to a series of QoS constraint of all communication users involved in the system, that is 
\begin{equation}
 \label{eq:problem1}
\begin{aligned}
(\mathcal{P}_1):
\begin{cases}
\max\limits_{\lbrace \pmb{w}_k \rbrace}&   P(\theta_0)\\
\textrm{s.t.}
 &  \Pr( \SINR_k \leq \gamma_k ) \leq p_k , \quad \forall k \\
 & \sum\limits_{k=1}^K \Tr(\pmb{W}_k) \leq 1, \\ 
 & \pmb{W}_k = \pmb{w}_k\pmb{w}_k^H  , \quad \forall k , \\
\end{cases}
\end{aligned}
\end{equation}
where the $\sum\limits_{k=1}^K \Tr(\pmb{W}_k) \leq 1$ is to reflect the available power budget\footnote{Indeed, one could easily replace the $1$ by a desired power budget, say $P_0$.} . Note that $(\mathcal{P}_1)$ is non-convex due to all the probabilistic constraints involved within the problem. One way of proceeding with the above problem is to derive the exact distribution function associated with the random variable $\SINR_k$, however the closed-form expression of the distribution function leads to a highly non-linear problem. To this end, we shall follow a series of relaxations to arrive at a tractable one.

\subsection{Semi-definite reformulation}
We start by re-expressing $(\mathcal{P}_1)$ as follows
 \begin{equation}
 \label{eq:problem1-again}
\begin{aligned}
(\mathcal{P}_1):
\begin{cases}
\max\limits_{\lbrace \pmb{w}_k \rbrace}&   P(\theta_0)\\
\textrm{s.t.}
 &  \Pr( 	\frac{ \frac{1}{\gamma_k} \tilde{\pmb{h}}_k ^T \pmb{W}_k \tilde{\pmb{h}}_k^*  } 
	     {\sum\limits_{\ell =1, \ell \neq k }^K \tilde{\pmb{h}}_k^T \pmb{W}_{\ell} \tilde{\pmb{h}}_k^* + \sigma_{c}^2} \leq 1 ) \leq p_k , \quad \forall k \\
 & \sum\limits_{k=1}^K \Tr(\pmb{W}_k) \leq 1, \\ 
 & \pmb{W}_k = \pmb{w}_k\pmb{w}_k^H  , \quad \forall k . \\
\end{cases}
\end{aligned}
\end{equation}
Through straightforward manipulations, one can show that the contribution of the $\ell^{th}$ beamformer on the $k^{th}$ communication channel could be expressed as 
\begin{equation}
\label{eq:hkWlhk}
	\tilde{\pmb{h}}_k^T \pmb{W}_{\ell} \tilde{\pmb{h}}_k^* 
	= {\pmb{h}}_k^T \pmb{W}_{\ell} {\pmb{h}}_k^* 
	+ \Delta{\pmb{h}}_k^T \pmb{W}_{\ell} \Delta{\pmb{h}}_k^* 
	+ 2\Real(\Delta{\pmb{h}}_k^T \pmb{W}_{\ell}{\pmb{h}}_k^*) .
\end{equation}
Using equation \eqref{eq:hkWlhk} in $(\mathcal{P}_1)$ appearing in equation \eqref{eq:problem1-again}, we get
\begin{equation}
\label{eq:problem1-linear-proba}
\begin{aligned}
\begin{cases}
\max\limits_{\lbrace \pmb{w}_k \rbrace}&   P(\theta_0)\\
\textrm{s.t.}
 &  
\Pr\big( \Delta{\pmb{h}}_k^T \bar{\pmb{W}}_{k} \Delta{\pmb{h}}_k^*+ 2\Real(\Delta{\pmb{h}}_k^T \bar{\pmb{W}}_{k}{\pmb{h}}_k^*) \leq \sigma_k^2  \big) \leq p_k  	 \\
 & \sum\limits_{k=1}^K \Tr(\pmb{W}_k) \leq 1, \\ 
   & \pmb{W}_k = \pmb{w}_k\pmb{w}_k^H  , \quad \forall k \\
   & \sigma_k^2 = \sigma_c^2   -  {\pmb{h}}_k^T \bar{\pmb{W}}_{k} {\pmb{h}}_k^* , \quad \forall k , \\
\end{cases}
\end{aligned}
\end{equation}
where $\bar{\pmb{W}}_{k}  = \frac{1}{\gamma_k} \pmb{W}_k - \sum\limits_{\ell = 1, \ell \neq k}^K \pmb{W}_k$. Notice now that we have moved all random quantities in the probability to the left-hand side of the inequality within the probability event. To standardize the gaussian vector $\Delta \pmb{h}_k$, we shall introduce normalization factors as follows, 
\begin{equation}
\label{eq:problem1-linear-proba-normalized}
\begin{aligned}
\begin{cases}
\max\limits_{\lbrace \pmb{w}_k \rbrace}&   P(\theta_0)\\
\textrm{s.t.}
 &  
\Pr\big(\pmb{z}_k^H \pmb{A}_k \pmb{z}_k+ 2\Real(\pmb{z}_k^H \pmb{b}_k) \leq \sigma_k^2  \big) \leq p_k  	 \\
 & \sum\limits_{k=1}^K \Tr(\pmb{W}_k) \leq 1, \\ 
   & \pmb{W}_k = \pmb{w}_k\pmb{w}_k^H  , \quad \forall k \\
   & \sigma_k^2 = \sigma_c^2   -  {\pmb{h}}_k^T \bar{\pmb{W}}_{k} {\pmb{h}}_k^* , \quad \forall k , \\
\end{cases}
\end{aligned}
\end{equation}
where $\pmb{z}_k =\frac{1}{\sigma_{\Delta \pmb{h}_k}} \Delta \pmb{h}_k^* $, $\pmb{A}_k = \sigma_{\Delta \pmb{h}_k}^2 \bar{\pmb{W}}_{k}$ and $\pmb{b}_k = \sigma_{\Delta \pmb{h}_k} \bar{\pmb{W}}_{k}{\pmb{h}}_k^*$, where $\pmb{z}_k \sim \mathcal{N}(\pmb{0},\pmb{I}_N)2$ Now, making use of \textit{Bernstien-type inequality} for quadratic-type Gaussian processes\cite{Bernstein-paper} similar to conventional robust MIMO beamforming \cite{berstien-traditional-01}, we can say that for any $\epsilon_k \geq 0$, we have the following $\Pr\Big(\pmb{z}_k^H \pmb{A}_k \pmb{z}_k + 2\Real(\pmb{z}_k^H \pmb{b}_k) \leq U_k  \Big)\leq \exp(-\epsilon_k)$, 

where $U_k = \Tr(\pmb{A}_k) - \sqrt{2\epsilon_k}\sqrt{\Vert \vect(\pmb{A}_k) \Vert^2 + 2\Vert \pmb{b}_k \Vert^2} - \epsilon_k \lambda_k^-$, where $\lambda_k^- = \max(\lambda_{\max}(-\pmb{A}_k),0)$. This is fulfilled by first adjusting $\epsilon_k = -\log(p_k)$ so as to guarantee the same outage probability level as in $(\mathcal{P}_1)$ and upper-bounding $\sigma_k^2$ by $U_k$, we can formulate an equivalent problem 
\begin{equation}
\label{eq:problem1-linear-proba-bernstien}
(\mathcal{P}_2):
\begin{aligned}
\begin{cases}
\max\limits_{\lbrace \pmb{w}_k \rbrace}&   P(\theta_0)\\
\textrm{s.t.}
 &  
\sigma_k^2 \leq U_k  	 \\
   & \sigma_k^2 = \sigma_c^2   -  {\pmb{h}}_k^T \bar{\pmb{W}}_{k} {\pmb{h}}_k^* , \quad \forall k \\
    & \sum\limits_{k=1}^K \Tr(\pmb{W}_k) \leq 1, \\ 
    & \pmb{W}_k = \pmb{w}_k\pmb{w}_k^H  , \quad \forall k \\
   & U_k = \Tr(\pmb{A}_k) - \sqrt{2\epsilon_k}c_k - \epsilon_k \lambda_k^-, \quad \forall k \\
   & c_k = \sqrt{\Vert \vect(\pmb{A}_k) \Vert^2 + 2\Vert \pmb{b}_k \Vert^2}, \quad \forall k .
\end{cases}
\end{aligned}
\end{equation}
Notice that problem $(\mathcal{P}_2)$ does not contain a probability event within its constraint. Furthermore, we replace the quadratic constraint $\pmb{W}_k = \pmb{w}_k\pmb{w}_k^H$ with an equivalent one as follows
\begin{equation}
\label{eq:problem1-linear-proba-bernstien-expressed-with-rank}
(\mathcal{P}_2):
\begin{aligned}
\begin{cases}
\max\limits_{\lbrace \pmb{W}_k \rbrace}&   P(\theta_0)\\
\textrm{s.t.}
 &  
\sigma_k^2 \leq U_k  	 \\
   & \sigma_k^2 = \sigma_c^2   -  {\pmb{h}}_k^T \bar{\pmb{W}}_{k} {\pmb{h}}_k^* , \quad \forall k \\
    & \pmb{W}_k \succeq \pmb{0}, \quad \rank(\pmb{W}_k) = 1  , \quad \forall k \\
     & \sum\limits_{k=1}^K \Tr(\pmb{W}_k) \leq 1, \\ 
   & U_k = \Tr(\pmb{A}_k) - \sqrt{2\epsilon_k}c_k - \epsilon_k \lambda_k^-, \quad \forall k \\
   & c_k = \sqrt{\Vert \vect(\pmb{A}_k) \Vert^2 + 2\Vert \pmb{b}_k \Vert^2}, \quad \forall k .
\end{cases}
\end{aligned}
\end{equation}
Next, we introduce $K$ non-negative slack variables, $\nu_1 \ldots \nu_K$, where $\nu_k$ aims at upper bounding the maximum eigenvalue of $-\pmb{A}_k$ as 
\begin{equation}
\label{eq:eigenConstraint}
	\nu_k  \pmb{I}_N  \succeq \lambda_k^- \pmb{I}_N \succeq - \pmb{A}_k .
\end{equation}
Introducing the inequality in equation \eqref{eq:eigenConstraint} to problem $(\mathcal{P}_2)$
\begin{equation}
\label{eq:problem1-linear-proba-bernstien-eigen}
(\mathcal{P}_3):
\begin{aligned}
\begin{cases}
\max\limits_{\lbrace \pmb{W}_k ,\nu_k,\mu_k \rbrace}&   P(\theta_0)\\
\textrm{s.t.}
 &  
\sigma_k^2 \leq U_k  	 \\
   & \sigma_k^2 = \sigma_c^2   -  {\pmb{h}}_k^T \bar{\pmb{W}}_{k} {\pmb{h}}_k^* , \quad \forall k \\
   & \nu_k  \pmb{I}_N  \succeq - \pmb{A}_k, \quad \forall k  \\
    & \pmb{W}_k \succeq \pmb{0}, \quad \rank(\pmb{W}_k) = 1  , \quad \forall k \\
     & \sum\limits_{k=1}^K \Tr(\pmb{W}_k) \leq 1, \\ 
   & L_k = \Tr(\pmb{A}_k) - \sqrt{2\epsilon_k}c_k - \epsilon_k \nu_k, \quad \forall k \\ 
   & U_k = \Tr(\pmb{A}_k) - \sqrt{2\epsilon_k}c_k - \epsilon_k \lambda_k^-, \quad \forall k \\
   & L_k \leq U_k, \quad \forall k  \\
   & c_k = \sqrt{\Vert \vect(\pmb{A}_k) \Vert^2 + 2\Vert \pmb{b}_k \Vert^2}, \quad \forall k \\
   & \nu_k \geq 0, \quad \forall k . \\
\end{cases}
\end{aligned}
\end{equation}
Now to resolve constraint $c_k = \sqrt{\Vert \vect(\pmb{A}_k) \Vert^2 + 2\Vert \pmb{b}_k \Vert^2}$, we will introduce $K$ additional non-negative slack variables, $\mu_1 \ldots \mu_K$ in the form of second-order cone (SOC) constraints as follows for all $k$ 
\begin{equation}
	\label{eq:socp-constraint}
	\mu_k \geq\sqrt{  \Vert  \vect(\pmb{A}_k) \Vert^2 + 2\Vert \pmb{b}_k \Vert^2} ,
\end{equation}
which further relaxes $(\mathcal{P}_3)$ as 
\begin{equation}
\label{eq:problem1-linear-proba-bernstien-socp-relax}
(\mathcal{P}_4):
\begin{aligned}
\begin{cases}
\max\limits_{\lbrace \pmb{W}_k ,\nu_k, \mu_k \rbrace}&   P(\theta_0)\\
\textrm{s.t.}
 &  
\sigma_k^2 \leq U_k  	 \\
   & \sigma_k^2 = \sigma_c^2   -  {\pmb{h}}_k^T \bar{\pmb{W}}_{k} {\pmb{h}}_k^* , \quad \forall k \\
   & \nu_k  \pmb{I}_N  \succeq - \pmb{A}_k, \quad \forall k  \\
    & \pmb{W}_k \succeq \pmb{0}, \quad \rank(\pmb{W}_k) = 1  , \quad \forall k \\
     & \sum\limits_{k=1}^K \Tr(\pmb{W}_k) \leq 1, \\ 
   & L_k = \Tr(\pmb{A}_k) - \sqrt{2\epsilon_k}\mu_k - \epsilon_k \nu_k, \quad \forall k \\ 
   & U_k = \Tr(\pmb{A}_k) - \sqrt{2\epsilon_k}c_k - \epsilon_k \lambda_k^-, \quad \forall k \\
   & L_k \leq U_k, \quad \forall k  \\
   & c_k = \sqrt{\Vert \vect(\pmb{A}_k) \Vert^2 + 2\Vert \pmb{b}_k \Vert^2}, \quad \forall k \\
   & \mu_k \geq c_k , \nu_k \geq 0,  \quad \forall k  \\
\end{cases}
\end{aligned}
\end{equation}
The SOC constraint in equation \eqref{eq:socp-constraint} could be re-written under a Schur complement type expression, viz.
\begin{equation}
	\mu_k \pmb{I}_{N+N^2}
	-
	\begin{bmatrix}
		\sqrt{2}\pmb{b}_k \\
		\vect(\pmb{A}_k)
	\end{bmatrix}
		\mu_k^{-1}
	\begin{bmatrix}
		\sqrt{2}\pmb{b}_k \\
		\vect(\pmb{A}_k)
	\end{bmatrix}^H 
	\succeq
	\pmb{0} ,
\end{equation}
which is also equivalent to 
\begin{equation}
\label{eq:Schur}
\pmb{Q}_k 
\eqdef
\left[
	\begin{array}{c|c}
		\mu_k  & 
	\begin{array}{cc}
		\sqrt{2}\pmb{b}_k^H &
		\vect(\pmb{A}_k)^H
	\end{array} \\
	\hline
	\begin{array}{cc}
		\sqrt{2}\pmb{b}_k \\
		\vect(\pmb{A}_k)
	\end{array} &
	\mu_k \pmb{I}_{N+N^2}
	\end{array}
	\right]
	\succeq
	\pmb{0},
	\quad
	\forall k .
\end{equation}
Injecting the above inequality within problem $(\mathcal{P}_4)$ we get 
\begin{equation}
\label{eq:problem1-linear-proba-bernstien-socp-relax}
(\mathcal{P}_5):
\begin{aligned}
\begin{cases}
\max\limits_{\Big\lbrace\substack{ \pmb{W}_k \\ \nu_k \\ \mu_k }\Big\rbrace}&   P(\theta_0)\\
\textrm{s.t.}
 &  
\sigma_k^2 \leq U_k  	 \\
   & \sigma_k^2 = \sigma_c^2   -  {\pmb{h}}_k^T \bar{\pmb{W}}_{k} {\pmb{h}}_k^* , \quad \forall k \\
    & \nu_k  \pmb{I}_N  \succeq - \pmb{A}_k, \quad \forall k  \\
    & \pmb{W}_k \succeq \pmb{0} , \quad \rank(\pmb{W}_k) = 1 , \quad \forall k \\
     & \sum\limits_{k=1}^K \Tr(\pmb{W}_k) \leq 1, \\ 
   & L_k = \Tr(\pmb{A}_k) - \sqrt{2\epsilon_k}\mu_k - \epsilon_k \nu_k, \quad \forall k \\ 
   & U_k = \Tr(\pmb{A}_k) - \sqrt{2\epsilon_k}c_k - \epsilon_k \lambda_k^-, \quad \forall k \\
   & L_k \leq U_k, \quad \forall k  \\
   & 	\pmb{Q}_k 
	\succeq
	\pmb{0}, \nu_k \geq 0, \quad \forall k .  \\
\end{cases}
\end{aligned}
\end{equation}  
Notice that the constraint $L_k \leq U_k$ is now a redundant constraint, due to the ones added from equations \eqref{eq:eigenConstraint} and \eqref{eq:Schur}. Therefore, we can eliminate $L_k \leq U_k$. Also, in order to completely remove the constraint $U_k = \Tr(\pmb{A}_k) - \sqrt{2\epsilon_k}c_k - \epsilon_k \lambda_k^-$, a reasonable modification would be to replace $\sigma_k^2 \leq U_k$ by $\sigma_k^2 \leq L_k$, since the latter guarantees the former.
\begin{equation}
\label{eq:problem1-linear-proba-bernstien-socp-relax}
(\mathcal{P}_6):
\begin{aligned}
\begin{cases}
\max\limits_{\Big\lbrace\substack{ \pmb{W}_k \\ \nu_k \\ \mu_k }\Big\rbrace}&   P(\theta_0)\\
\textrm{s.t.}
 &  
\sigma_k^2 \leq \Tr(\pmb{A}_k) - \sqrt{2\epsilon_k}\mu_k - \epsilon_k \nu_k  	 \\
   & \sigma_k^2 = \sigma_c^2   -  {\pmb{h}}_k^T \bar{\pmb{W}}_{k} {\pmb{h}}_k^* , \quad \forall k \\
    & \nu_k  \pmb{I}_N  \succeq - \pmb{A}_k, \quad \forall k  \\
    & \pmb{W}_k \succeq \pmb{0}, \quad \rank(\pmb{W}_k) = 1  , \quad \forall k \\
     & \sum\limits_{k=1}^K \Tr(\pmb{W}_k) \leq 1, \\ 
   & 	\pmb{Q}_k \succeq \pmb{0}, \nu_k \geq 0, \quad \forall k . \\
\end{cases}
\end{aligned}
\end{equation}
The problem in $(\mathcal{P}_6)$ is still non-convex due to the $\rank$-1 constraint. A classical remedy to convexify the problem is to drop the $\rank$-1 constraint, which finally leads to problem $(\mathcal{P}_7)$ as follows
\begin{equation}
\label{eq:problem1-linear-proba-bernstien-socp-relax-final}
(\mathcal{P}_7):
\begin{aligned}
\begin{cases}
\max\limits_{\Big\lbrace\substack{ \pmb{W}_k \\ \nu_k \\ \mu_k }\Big\rbrace}&   P(\theta_0)\\
\textrm{s.t.}
 &  
\sigma_k^2 \leq \Tr(\pmb{A}_k) - \sqrt{2\epsilon_k}\mu_k - \epsilon_k \nu_k  	 \\
   & \sigma_k^2 = \sigma_c^2   -  {\pmb{h}}_k^T \bar{\pmb{W}}_{k} {\pmb{h}}_k^* , \quad \forall k \\
    & \nu_k  \pmb{I}_N  \succeq - \pmb{A}_k, \quad \forall k  \\
    & \pmb{W}_k \succeq \pmb{0}, \quad \forall k \\
     & \sum\limits_{k=1}^K \Tr(\pmb{W}_k) \leq 1, \\ 
   & 	\pmb{Q}_k \succeq \pmb{0}, \nu_k \geq 0, \quad \forall k . \\
\end{cases}
\end{aligned}
\end{equation}
Finally, we now have a convex optimization problem at our disposal, that could be easily solved using any convex optimization solver, such as the CVX toolbox \cite{cvx}.

\subsection{Rank-one beamforming}
In practical PHY layers, beamforming should take place through $\rank$-one solutions, i.e. vectors of the form $\pmb{w}_k\pmb{w}_k^H = \pmb{W}_k$. Otherwise, sub-optimal post-processing methods are needed to provide $\rank$-one approximations of $\pmb{W}_k$. In the literature, the most common approximations are Gaussian randomization \cite{GaussianRandomization, Noma-outage} and principle component analysis \cite{PCA} on $\pmb{W}_k$. However, in some particular scenarios, the $\rank$-one relaxation does not introduce any sort of suboptimality due to the special structure of the problem. For example, the relaxation in  \cite{ottersten-paulraj} has been proven to be exact and thus always gives $\rank$-1 optimal solution in the context of collaborative relay beamforming. Another instance appears in \cite{rankone-jsac} for joint sensing and communication design with low-PAPR waveform constraints. To this extent, we introduce the following theorem: \\\\
\textbf{Theorem 1 (Rank-one Optimality): } \textit{Consider the convex optimization problem 
$(\mathcal{P}_7)$ given in equation \eqref{eq:problem1-linear-proba-bernstien-socp-relax-final}. Then, for all $k = 1 \ldots K$, the solution of $(\mathcal{P}_7)$ satisfies }
\begin{equation}
	\rank(\pmb{W}_k^{\opt}) = 1 .
\end{equation}
\textbf{\emph{Proof:}}  See Appendix \ref{rankone}.

A $\rank$-1 proof appears in the context of resource allocation for intelligent reflective surfaces assisted full-duplex cognitive radio systems \cite{rank1IRS}. Now, that we have a guarantee of $\rank-$one solutions, an eigenvalue decomposition per $\pmb{W}_k^{\opt}$ is done to recover $\pmb{w}_k^{\opt}$.
\subsection{Single-user closed-form solutions}

\begin{figure*}[!t]
\centering
\subfloat[Case where $\Lambda \leq \vert \pmb{h}^T \pmb{a}^*(\theta_0) \vert^2$]{\includegraphics[]{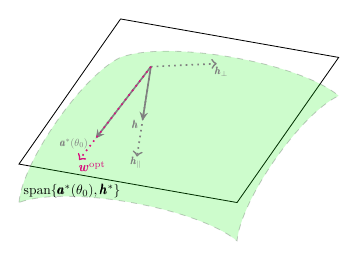} 
\label{fig:subspace-01}}
\hfil
\subfloat[Case where $\Lambda > \vert \pmb{h}^T \pmb{a}^*(\theta_0) \vert^2$]{\includegraphics[]{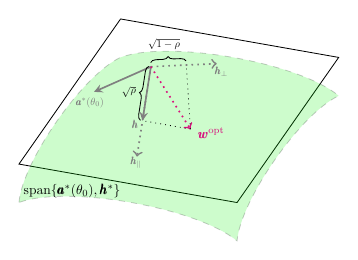}  
\label{fig:subspace-02}}
\caption{Graphical interpretation of the DFRC beamforming weight vector, as a function of steering vector and mean channel response.}
\label{fig:subspaces-in-action}
\end{figure*}

In the single user case, we bring special insights to the joint sensing and communication optimization problem in hand. To this end, and for $K=1$, the problem appearing in $(\mathcal{P}_6)$ boils down to
\begin{equation}
\label{eq:problem1-linear-proba-bernstien-socp-SU}
(\mathcal{P}_6^{\text{SU}}):
\begin{aligned}
\begin{cases}
\max\limits_{\lbrace \pmb{w} ,\nu, \mu \rbrace}&   
\big\vert \pmb{a}^T(\theta_0) \pmb{w} \big\vert^2 \\
\textrm{s.t.}
 &  
\sigma_c^2   -  \frac{1}{\gamma} \vert {\pmb{h}}^T \pmb{w} \vert^2 \leq \frac{\sigma_{\Delta}^2 }{\gamma}\Vert \pmb{w} \Vert^2  - \sqrt{2\epsilon}\mu - \epsilon \nu  	 \\
    & \nu  \pmb{I}_N  \succeq - \frac{\sigma_{\Delta}^2 }{\gamma} \pmb{w}\pmb{w}^H  \\
     & \Vert \pmb{w} \Vert^2 \leq 1 \\ 
   & 	\mu \geq  \sqrt{\Vert \vect(\pmb{A}) \Vert^2 + 2\Vert \pmb{b} \Vert^2}\\
   & \nu \geq 0 , \\
\end{cases}
\end{aligned}
\end{equation}
where we have denoted, for sake of simplicity, the single user channel as $\pmb{h} \eqdef \pmb{h}_1$, the $\SINR$ threshold as $\gamma \eqdef \gamma_1$, the channel error variance as $\sigma_{\Delta} \eqdef \sigma^2_{\Delta \pmb{h}_1}$, the slack variables $\nu_1 \eqdef \nu$, $\mu_1 \eqdef \mu $, the  $\epsilon_1 \eqdef \epsilon$ and finally the beamforming vector $\pmb{w}_1 \eqdef \pmb{w}$. Note that the semi-definite constraint $\nu  \pmb{I}_N  \succeq - \frac{\sigma_{\Delta}^2 }{\gamma} \pmb{w}\pmb{w}^H$ along with the non-negativity constraint on $\nu$ defines the region $\nu \geq 0 \geq  -\frac{\sigma_{\Delta}^2 }{\gamma} \Vert \pmb{w} \Vert^2$. 
This region along with the identities $\Vert \vect(\pmb{A}) \Vert^2 = \Tr(\pmb{A}^H \pmb{A}) = \frac{\sigma_{\Delta}^4}{\gamma^2} \Vert \pmb{w} \Vert^4$ and $\Vert \pmb{b} \Vert^2 = \frac{\sigma_{\Delta}^2}{\gamma^2}\big\vert \pmb{h}^T \pmb{w} \big\vert^2 \Vert\pmb{w}\Vert^2$ are used to upper bound the first constraint appearing in problem \eqref{eq:problem1-linear-proba-bernstien-socp-SU} to get the problem below, which no longer depends on $\nu$ and $\mu$, i.e.
\begin{equation}
\label{eq:problem1-linear-proba-bernstien-socp-SU2}
(\mathcal{P}_6^{\text{SU}}):
\begin{aligned}
\begin{cases}
\max\limits_{ \pmb{w}}&   
\big\vert \pmb{a}^T(\theta_0) \pmb{w} \big\vert^2 \\
\textrm{s.t.}
 &  \gamma\sigma_c^2   - f_{\epsilon}(\pmb{w})  \Vert \pmb{w} \Vert^2  \leq    \vert {\pmb{h}}^T \pmb{w}  \vert^2 \\
     & \Vert \pmb{w} \Vert^2 \leq 1, \\ 
\end{cases}
\end{aligned}
\end{equation}
where $f_{\epsilon}(\pmb{w}) = \sigma_{\Delta}^2 - \sqrt{2\epsilon} \sigma_{\Delta}\sqrt{\sigma_{\Delta}^2 + 2 \frac{\big\vert \pmb{h}^T \pmb{w} \big\vert^2}{\Vert\pmb{w}\Vert^2} } $. We now present the next theorem:

\textbf{Theorem 2 (Single-User BFing in Closed-Form, $\epsilon \rightarrow 0 $):} \textit{Let $K = 1$, the optimal solution to problem $(\mathcal{P}_6^{\text{SU}})$ when $\epsilon = 0$, is given as follows}
\begin{equation}
\label{eq:closed-form-single-user-for-epsilon-zero}
	\pmb{w}^{\opt}
	=
	\begin{cases}
		\frac{\pmb{a}^*(\theta_0)}{\Vert \pmb{a}(\theta_0) \Vert},   \qquad\text{if } N(\gamma\sigma_c^2   - \sigma_{\Delta}^2 ) \leq     
 \vert {\pmb{h}}^T \pmb{a}^*(\theta_0) \vert^2 \\
	\sqrt{\rho}\frac{\pmb{h}_{\parallel}^H \pmb{a}^*(\theta_0)}{\Vert \pmb{h}_{\parallel}^H \pmb{a}^*(\theta_0) \Vert} \pmb{h}_{\parallel} + 
	\sqrt{1-\rho}\frac{\pmb{h}_{\perp}^H \pmb{a}^*(\theta_0)}{\Vert \pmb{h}_{\perp}	^H \pmb{a}^*(\theta_0) \Vert}\pmb{h}_{\perp} , \qquad   \text{else}
	\end{cases}
\end{equation}
\textit{given the following feasibility condition}
\begin{equation}
 0 \leq \rho \eqdef \frac{\gamma \sigma_c^2 - \sigma_{\Delta}^2 }{\Vert \pmb{h} \Vert^2} \leq 1 .
\end{equation}
\textbf{\emph{Proof:}}  See Appendix \ref{app:su-solution}.

The closed-form solution in equation \eqref{eq:closed-form-single-user-for-epsilon-zero} brings many useful insights. When the channel $\pmb{h}$ and the steering vector towards the look direction, i.e. $\pmb{a}(\theta_0)$, are highly correlated, then the solution is the Bartlett beamformer. In other words, the efforts spent on maximizing the $\SINR$ are obtained via the radar metric. Note that the threshold that determines the degree of correlation between the channel and the steering vector in the look direction is $N(\gamma\sigma_c^2   - \sigma_{\Delta}^2 )$. From here, we can see that the given $\gamma$ has an impact on deciding the radar-communication tradeoff. With increasing $\gamma$, we are enforcing a strong correlation between the channel and the steering vectors to return a full radar solution. On the other hand, if low correlation exists, then both the channel and the steering vectors contribute to the solution. Specifically, and from a communication perspective, an extreme case is realized upon adjusting $\gamma$ so that $\rho = 1$, where the optimal beamforming vector is now a matched filter solution. Interestingly, the feasibility of the problem is expressed as the maximum achievable average-$\SINR$ defined in equation \eqref{eq:max-achieve}. We can further generalize \textbf{Theorem 2} as follows\\
\textbf{Theorem 3 (Single-User BFing in Closed-Form):} \textit{Let $K = 1$, the optimal solution to problem $(\mathcal{P}_6^{\text{SU}})$, is given as follows}
\begin{equation}
	\pmb{w}^{\opt}
	=
	\begin{cases}
		\frac{\pmb{a}^*(\theta_0)}{\Vert \pmb{a}(\theta_0) \Vert},   \qquad\text{if } \Lambda \leq     
 \vert {\pmb{h}}^T \pmb{a}^*(\theta_0) \vert^2 \\
	\sqrt{\rho}\frac{\pmb{h}_{\parallel}^H \pmb{a}^*(\theta_0)}{\Vert \pmb{h}_{\parallel}^H \pmb{a}^*(\theta_0) \Vert} \pmb{h}_{\parallel} + 
	\sqrt{1-\rho}\frac{\pmb{h}_{\perp}^H \pmb{a}^*(\theta_0)}{\Vert \pmb{h}_{\perp}	^H \pmb{a}^*(\theta_0) \Vert}\pmb{h}_{\perp} , \qquad   \text{else}
	\end{cases}
\end{equation}
\textit{where} $\Lambda = N(\gamma\sigma_c^2   - \sigma_{\Delta}^2 + \sqrt{2\epsilon}\sigma_{\Delta}\sqrt{\sigma_{\Delta}^2 + 2 \frac{\vert {\pmb{h}}^T \pmb{a}^*(\theta_0) \vert^2}{N}} )$
\textit{and $\sqrt{\rho}$ is a root of the following equation in the interval $[0,1]$}
\begin{equation}
	g(x) = x^2 \Vert \pmb{h} \Vert^2 - (\gamma \sigma_c^2 - \sigma_{\Delta}^2)-\sigma_{\Delta}\sqrt{2\epsilon}\sqrt{\sigma_{\Delta}^2 + x^2 \Vert \pmb{h} \Vert^2} . 
\end{equation}
The proof of the above theorem follows similar steps as in the proof found in Appendix \ref{app:su-solution}. Note that when $\epsilon \rightarrow 0$, the expressions found in \textbf{Theorem 3} coincide with that in \textbf{Theorem 2}.

Note that \textbf{Theorem 3} reveals the impact of outage probability rate $p$, or equivalently $\epsilon$. In particular, the more we restrict a tighter outage $\SINR$ failure, the larger $\Lambda$ is. To summarize, both $\gamma$ and $p$ have a direct influence on the radar-communication tradeoffs as $\Lambda$ scales as $\mathcal{O}(\gamma)$ and $\mathcal{O}(\epsilon^\frac{1}{2})$. To illustrate this, a graphical interpretation is given in Fig. \ref{fig:subspaces-in-action}.

\section{Simulation Results}
\label{sec:simulations}
In this section, various simulation results are conducted to illustrate the performance and trade-offs achieved with the proposed beamforming design. Through all simulations, we assume that the channel error variations are equal, namely $\sigma_{\Delta h_1} = \ldots = \sigma_{\Delta h_K} = \sigma_{\Delta}$. Unless otherwise stated, we set $\sigma_{\Delta} = 0.1$. Likewise, we set all outage parameters per communication user to be the same, i.e. $p_1 = \ldots = p_K = p$ and $\gamma_1 = \ldots = \gamma_K = \gamma$. Simulation results are realized through Monte Carlo averaging, where each point generated is averaged over $10^3$ Monte Carlo trials. The channels $\pmb{h}_k$ and the imperfect CSI $\Delta \pmb{h}_k$ are drawn from a complex Gaussian distribution with zero mean. We have also used CVX as a solver \cite{cvx} to generate the beamforming solutions of problem $(\mathcal{P}_7)$. A uniform linear array spaced at half a wavelength apart has been considered.

\begin{figure}[!t]
\centering
\includegraphics[width=3.5in]{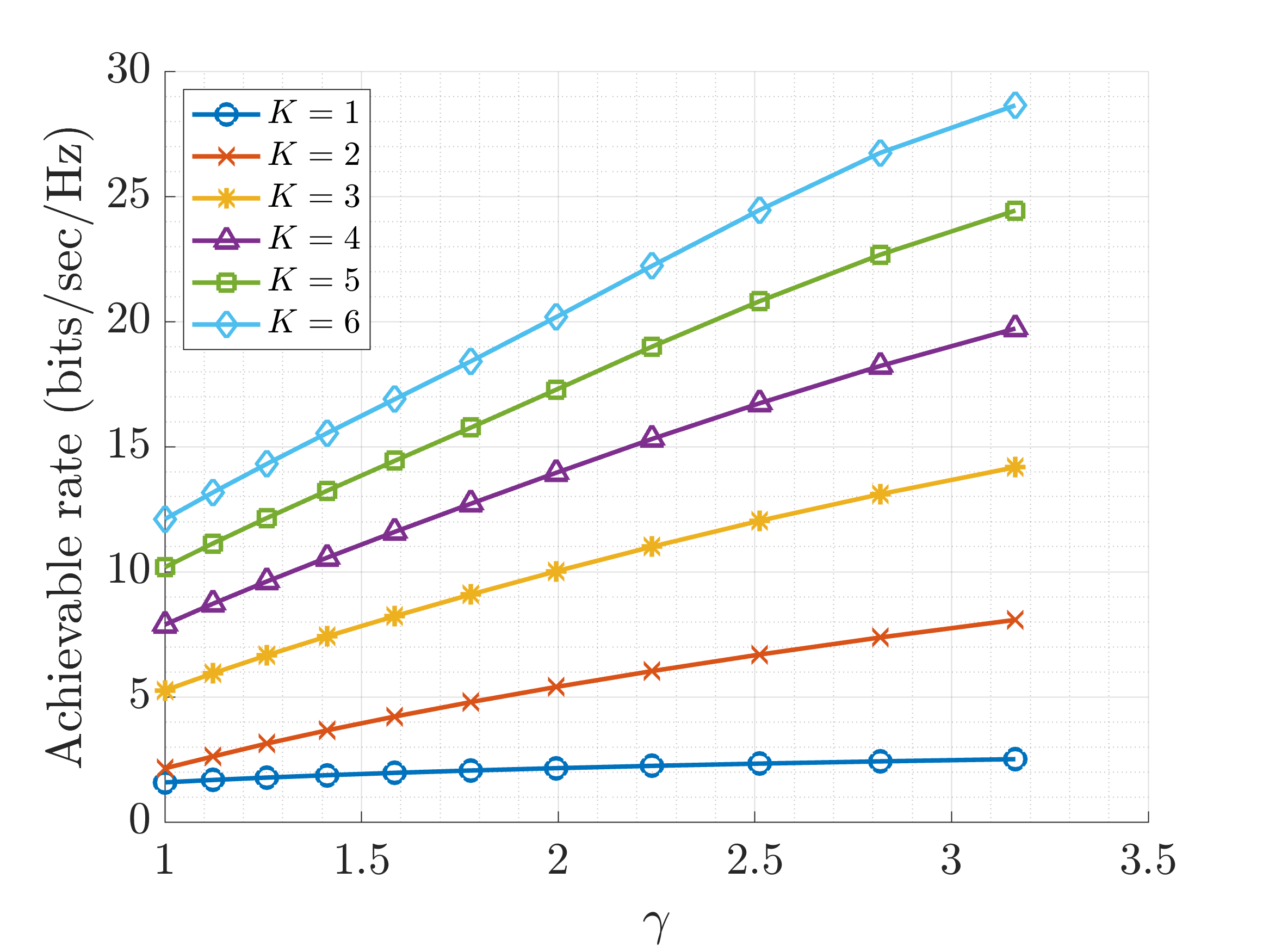}
\caption{Achievable rates as a function of $\gamma$ for different values of $K$ at $N = 10$ antennas and $p = 0.1$.}
\label{fig:rate-vs-gamma}
\end{figure}

\begin{figure}[!t]
\centering
\includegraphics[width=3.5in]{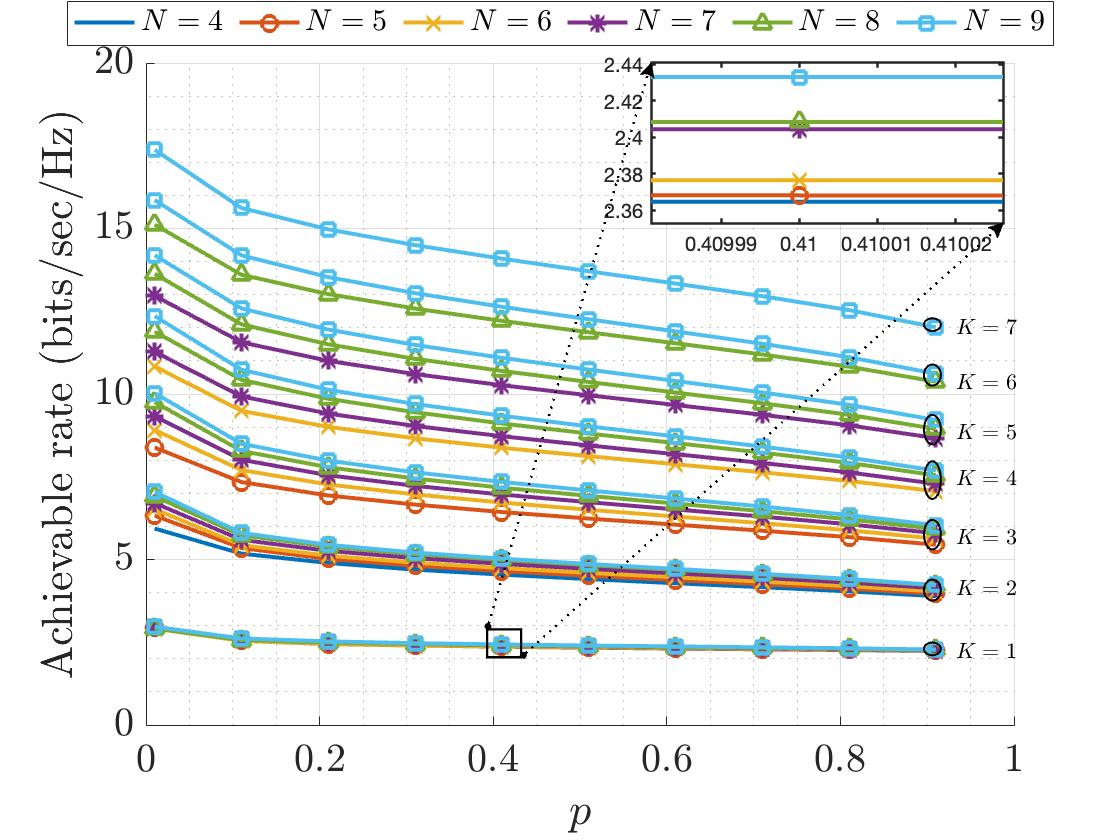}
\caption{Achievable rates as a function of $p$ for different values of $N$ and $K$.}
\label{fig:rate-vs-p}
\end{figure}

\subsection{Influence of $\gamma$ on the achievable rate}
For each Monte-Carlo simulation, we compute the $\SINR$ for each communication user, then compute the achieved rate using \eqref{eq:rate-equation}. First, we calculate the $\SINR$ for communication user $k$, at the $p^{th}$ Monte-Carlo simulations, which is defined as
\begin{equation}
	\SINR_k^{(p)}
	=
	\frac{ (\tilde{\pmb{h}}_k^{(p)}) ^T \pmb{w}_k^{\tt{sol}}(\pmb{w}_k^{\tt{sol}})^H (\tilde{\pmb{h}}_k^{(p)})^*  } 
	     {\sum\limits_{\ell =1, \ell \neq k }^K (\tilde{\pmb{h}}_k^{(p)})^T \pmb{w}_k^{\tt{sol}}(\pmb{w}_k^{\tt{sol}})^H (\tilde{\pmb{h}}_k^{(p)})^* + \sigma_{c}^2},	
\end{equation}
where $\pmb{w}_k^{\tt{sol}}$ is the beamforming vector computed by problem $(\mathcal{P}_7)$ and $\tilde{\pmb{h}}_k^{(p)}$ is the actual channel vector at the $p^{th}$ Monte-Carlo simulations towards the  $k^{th}$ communication user. Then, we evaluate the achieved rate as $R_k^{(p)} = \log_2 ( 1 + \SINR_k^{(p)}) $. If outage occurs, then $R_k^{(p)}$ at this time is outaged and set to $0$. Averaging over different Monte-Carlo simulations, we get $\bar{R}_k = \frac{1}{P} \sum\limits_{p=1}^P  R_k^{(p)}$, then we sum the rate over all users to get the total achievable rate.
We start the simulation section by studying the total achievable rate, when $\gamma$ is the varying factor, in Fig. \ref{fig:rate-vs-gamma}. We set $p = 0.1$ and $\theta_0 = 30^{\circ}$. In terms of rate achievability, and as expected, we observe a rising trend with increasing $\gamma$ or increasing number of users. For sake of argument, let us first focus on the case of $K = 2$, we can see that at $\gamma = 1\dB$, the achieved rate is very close to the threshold, i.e. $2 \log_2 ( 1 + \gamma) \simeq 2.17 \bpsph $ . On the other hand, notice that at $\gamma = 2\dB$, we are expecting a total rate of at least $2.35 \bpsph$, however, the actual achieved rate is $5.39\bpsph$. Even though users are subject to imperfect CSI, we observe a rising trend between the total achievable rate and the one being probed for, with increasing number of users, as well. As we will see in simulations below, this trend could be controlled through the outage probability $p$.
\subsection{Influence of $p$ on the achievable rate}
In the simulations shown in Fig. \ref{fig:rate-vs-p}, we aim at studying the impact of $p$ on the total achievable rate with different number of transmit antennas and communication users. It is clear that increasing $p$ leads to a degradation in the total achievable rate. In particular, a breakout region is observed around $p = 0.1$, where beyond this value, a stable slope contributes in the decrease of the total achievable rate. On another note, the degrees of freedom (DoF) available accounts for additional improvement in terms of $\SINR$, due to additional nulling of the interference terms; therefore, an increase in the total achievable rate. As an example, for the case of $K=2$, doubling the number of antennas contribute to an increase of roughly $0.3\bpsph$, whereas the same rate increase could be attained by adding only $1$ antenna, when $K = 5$.

\begin{figure*}[!t]
	\centering
	\includegraphics[width=1\linewidth]{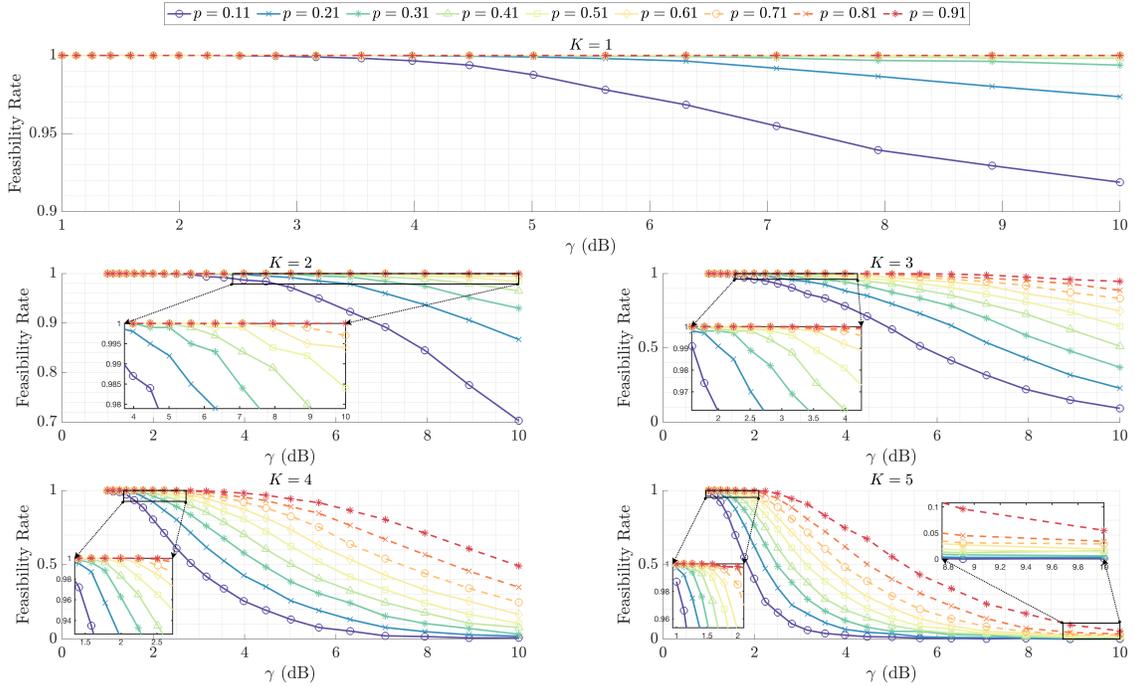}
  \caption{Feasibility rate behavior with varying $\gamma$ at $N = 6$ for different number of communication users at a target located at $\theta_0 = 30^\circ$.}
	\label{fig:feasibility-rate}
\end{figure*}

\subsection{Feasibility rate}
Another aspect we intend to shed light on is the $\feasibilityrate$. A solution is deemed infeasible when either one of the constraints of problem $\mathcal{P}_7$ is violated or when a solution is not available; therefore, a reasonable way to simulate is the $\feasibilityrate=
	\frac{\# \text{ of feasible solutions}}{\# \text{ of cases}}$.
Setting the number of transmit antennas to $N = 6$, we aim at concluding some allowable $\SINR$ thresholds with different number of communication users. As expected, the larger $\gamma$ is, the tighter is the feasibility region attained for our joint beamforming design, for any $p$ and $K$. Nonetheless, another nice interpretation of $p$ is its direct impact on the feasibility region. Said differently, a larger $p$ dilates the feasibility region, while sacrificing lower achievable rate, according to the previous simulations in Fig. \ref{fig:rate-vs-p}. For example, for the case of $K = 2$ communication users, our beamforming design is $99\%$ capable of returning a feasible solution for an $\SINR$ target of $\gamma = 3.8\dB$. At this feasibility level, one could increase the target by roughly $1.4\dB$, just by incrementing the outage rate by $0.1$, as depicted in Fig. \ref{fig:feasibility-rate} for $K=2$. Furthermore, as $K$ increases, it is evident that expected $\SINR$ target levels should shrink down, due to the unit power budget constraint. For example, at a feasibility rate of $99\%$, we see that we can target $\gamma = 1.6\dB$ for $K = 3$ users, compared to $\gamma = 1.25\dB$ for $K = 4$. 

 \subsection{$\ISMR$-rate trade-off}
 So far, we have only addressed the communication performance, without dedicating any attention to radar metrics. To include the impact of radar metrics, we discuss different radar-communication tradeoffs, with emphasis on two well-known radar metrics, namely, the integrated-sidelobe-to-mainlobe-ratio\footnote{Even though our beamforming design does not explicitly optimize $\ISMR$ terms, it does make absolute sense to include this metric in our analysis route. This is due to the fact that our cost function maximizes the output of the Bartlett beamformer in the look direction, $\theta_0$. Therefore, to measure this mainlobe power, the sidelobe power should be taken as reference.} ($\ISMR$) and the probability of detection. We start with the former.
 
\begin{figure}[!t]
\centering
\includegraphics[width=3.5in]{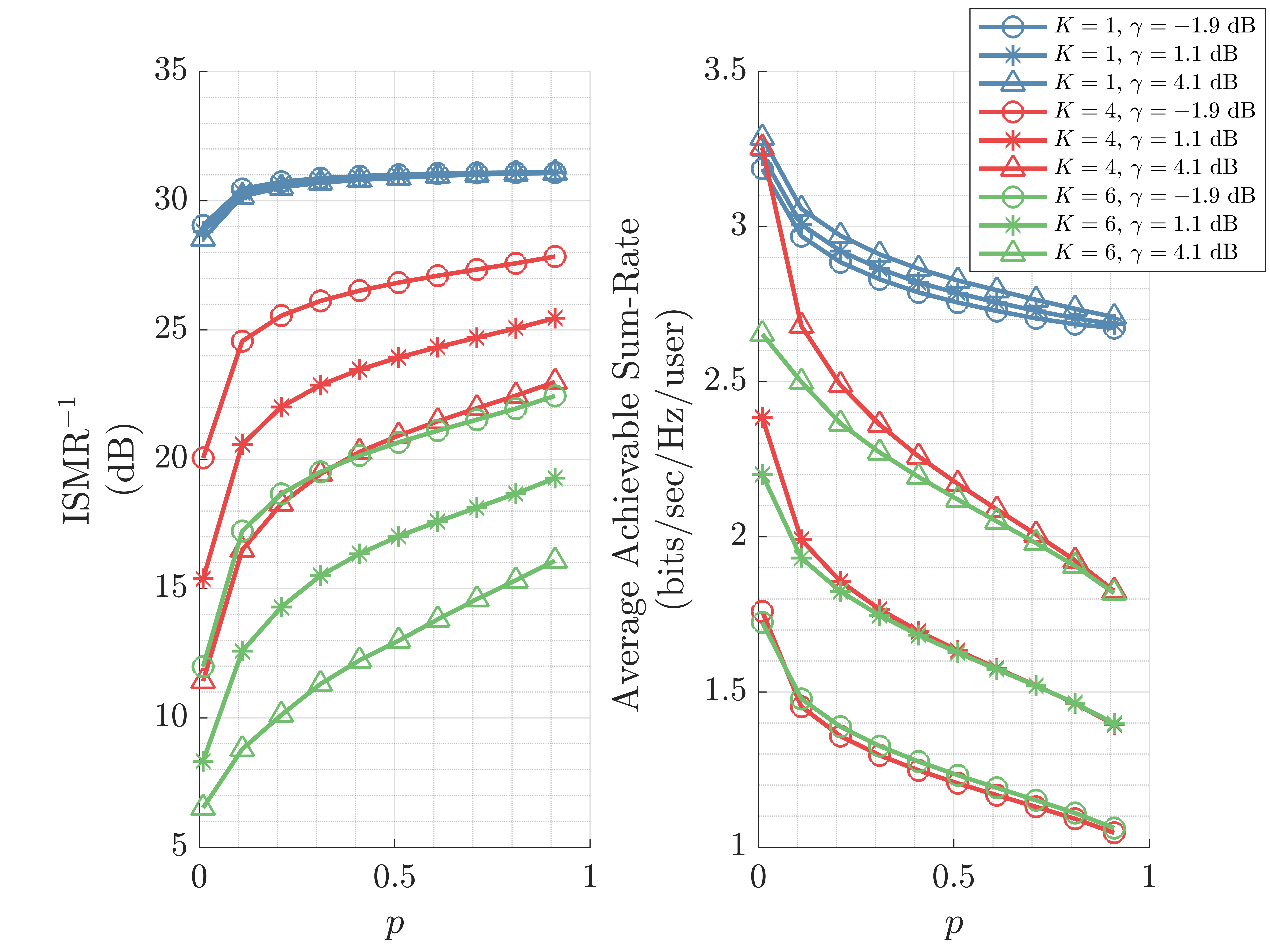}
\caption{Behavior of the average achievable sum rate per user and the $\ISMR$ for $N = 7$.}
\label{fig:ISMR-rate-tradeoff-01}
\end{figure}

The $\ISMR$ metric measures the energy in the sidelobe relative to the mainlobe and is defined as follows $\ISMR = \frac{\int_{\Theta_s} P(\theta) \ d\theta}{\int_{\Theta_m} P(\theta) \ d\theta} =  \frac{ \Tr( \pmb{W}\pmb{W}^H \pmb{A}_s) }{ \Tr( \pmb{W}\pmb{W}^H \pmb{A}_m)}$, 
  where $\Theta_m$ is the mainlobe region and $\Theta_s$ is the sidelobe region and $\pmb{A}_{s/m} = \int_{\Theta_{s/m}} \pmb{a}(\theta)\pmb{a}^H(\theta) \ d\theta $, whose expression for uniform linear arrays is given in \cite{LobeMatrices}. Note that the $\ISMR$ computation requires specification of the mainlobe and sidelobe supports. We have specified the mainlobe and sidelobe regions to be $\Theta_m = [\theta_0 \pm \frac{\Delta}{2}]$ and $\Theta_s = [-\frac{\pi}{2},\theta_0 - \frac{\Delta}{2}] \cup [\theta_0 + \frac{\Delta}{2},\frac{\pi}{2}]$, where we have set $\Delta = 20^\circ$. Moreover, its inverse, $\ISMR^{-1}$, reports the amount of energy concentrated in the mainlobe relative to that of the sidelobe. According to Fig. \ref{fig:ISMR-rate-tradeoff-01} and Fig. \ref{fig:ISMR-rate-tradeoff-02}, it is clear that a trade-off between the average communication rate and the radar $\ISMR$ performance exists. In particular, for a fixed $\ISMR^{-1}$ level, the average achievable rate increases with either an increase in $\gamma$ or a decrease in number of communication users, $K$. For example, fixing $\ISMR^{-1} = 22\dB$, one could increase the average achievable rate by roughly $0.33 \bpsph$ either by increasing $\gamma$ by a factor of $3\dB$, or by operating with $2$ less communication users, or by increasing the DoFs available. Therefore, both $\gamma$ and $p$ have a direct influence on the radar-communication tradeoff.

\subsection{Probability of detection and rate trade-off }
\begin{figure}[!t]
\centering
\includegraphics[width=3.5in]{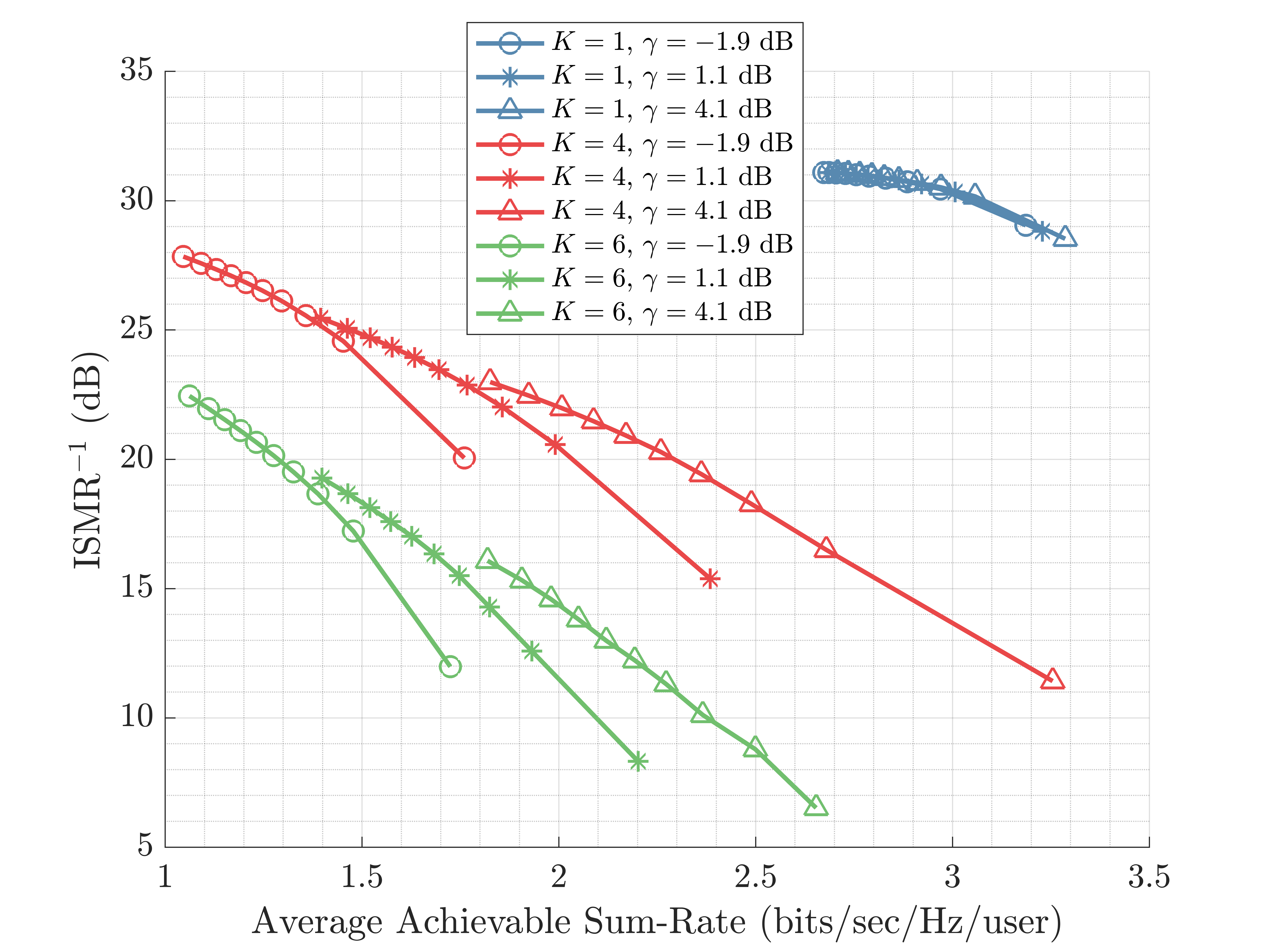}
\caption{Trade-off between the average achievable sum rate per user and the beampattern's $\ISMR$ for $N = 7$.}
\label{fig:ISMR-rate-tradeoff-02}
\end{figure}

Another tradeoff we can study is the probability of detection $P_D$, vs the average achievable sum rate. We calculate the probability of detection based on equation \eqref{eq:probability-of-detection}. The radar $\SNR$ is defined as $\SNR_r$ and is fixed to $1\dB$. We set $N = 5$ antennas. The probability of false-alarm for radar is $P_{\text{FA}} = 10^{-4}$. To show the superiority of our design, we compare to the beamforming design in \cite{toward-dfrc}. Referring to Fig. \ref{fig:PD-rate-tradeoff}, it is clear that a trade-off, similar to that in Fig. \ref{fig:ISMR-rate-tradeoff-02}, exists between the detection performance of the radar and the communication rate. In other words, for a fixed $P_D$, the rate increases with a decrease in number of users. Second, it should be noted that our design presents better tradeoffs in both, probability of detection and average achievable rate, especially when $K$ grows large. For example, with $K=4$ and an average achievable sum-rate of $4.78 \bpsphpu$, we have that $P_D \simeq 0.65$, when the design in \cite{toward-dfrc} is adopted; compared to $P_D \simeq 0.99$ using the proposed design, herein. This gain could be explained as follows: The design in \cite{toward-dfrc} is a two-step procedure. In the first step, the desired radar-only waveform is obtained through the desired covariance matrix, then in the second step, a weighted optimization problem is formulated that trades off a similarity constraint on the radar waveform and the MUI (Multi-user Interference). This decoupling of problems could introduce some sub-optimality to the joint radar-communication designs. Another aspect is the inherit modeling of random channel behavior in our design. On the other hand, our design cannot account for a desired beampattern, as opposed to that in \cite{toward-dfrc}. In other words, the proposed design does not offer any control over beamwidths, sidelobe rejection ratios, etc. This is due to the fact that we maximize the power in the desired look direction, without explicit indication of the aforementioned radar properties.

\begin{figure}[!t]
\centering
\includegraphics[width=3.5in]{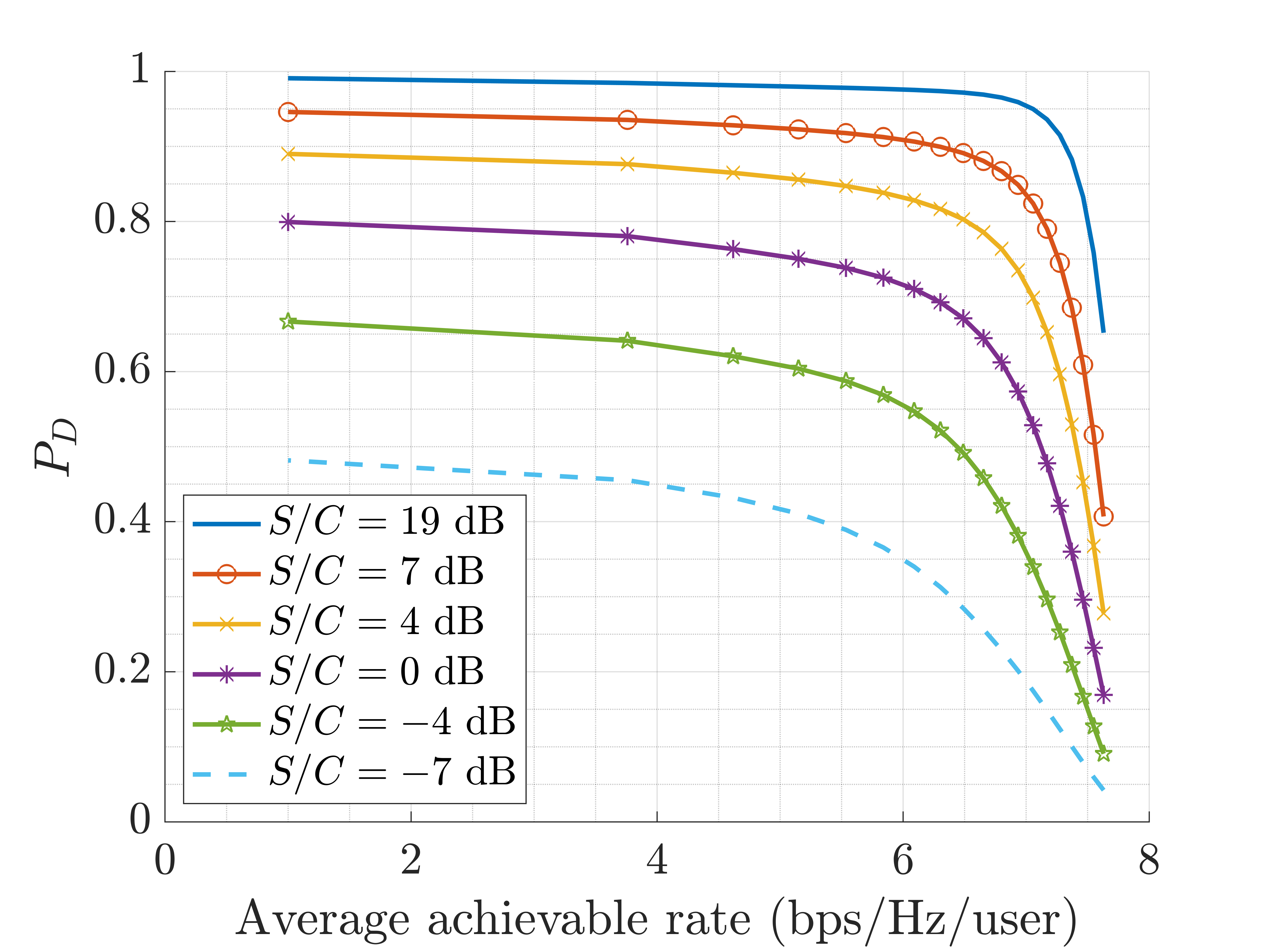}
\caption{The impact of the clutter on the tradeoff between communications and radar for the proposed beamforming design.}
\label{fig:clutter-tradeoff}
\end{figure}

To study the impact of the clutter on the trade-off between the probability of radar detection, we vary the signal-to-clutter ratio, denoted as $\frac{S}{C}$, and plot the $P_D$ versus average achievable rate curves obtained per $\frac{S}{C}$ value in Fig. \ref{fig:clutter-tradeoff}. As expected, a higher $\frac{S}{C}$ value contributes to a higher $P_D$ at the same communication rate. For example, given that $\frac{S}{C} \leq 7 \dB$ and at a fixed rate of $6 \bpsphpu$, we observe that an increase of around $3\dB$ in $\frac{S}{C}$ leads to an increase in $P_D$ in the order of $0.2$. Therefore, we conclude that the clutter has a direct consequence on the trade-off curves achieved by the proposed beamforming design.

\begin{figure}[!t]
  \includestandalone[width=0.5\textwidth]{./figures/job14}
  \caption{Trade-off between the average achievable sum rate per user and the radar detection
probability for $N = 5$ at receive $\SNR_r = 1 \dB$, and $P_{\text{FA}} = 10^{-4}$.}
  \label{fig:PD-rate-tradeoff}
\end{figure}

\subsection{Beampattern trade-offs}
\begin{figure*}[!t]
\centering
\subfloat[Target at $\theta_0 = 30^{\circ}$]{\includegraphics[height=2.25in,width=3.15in]{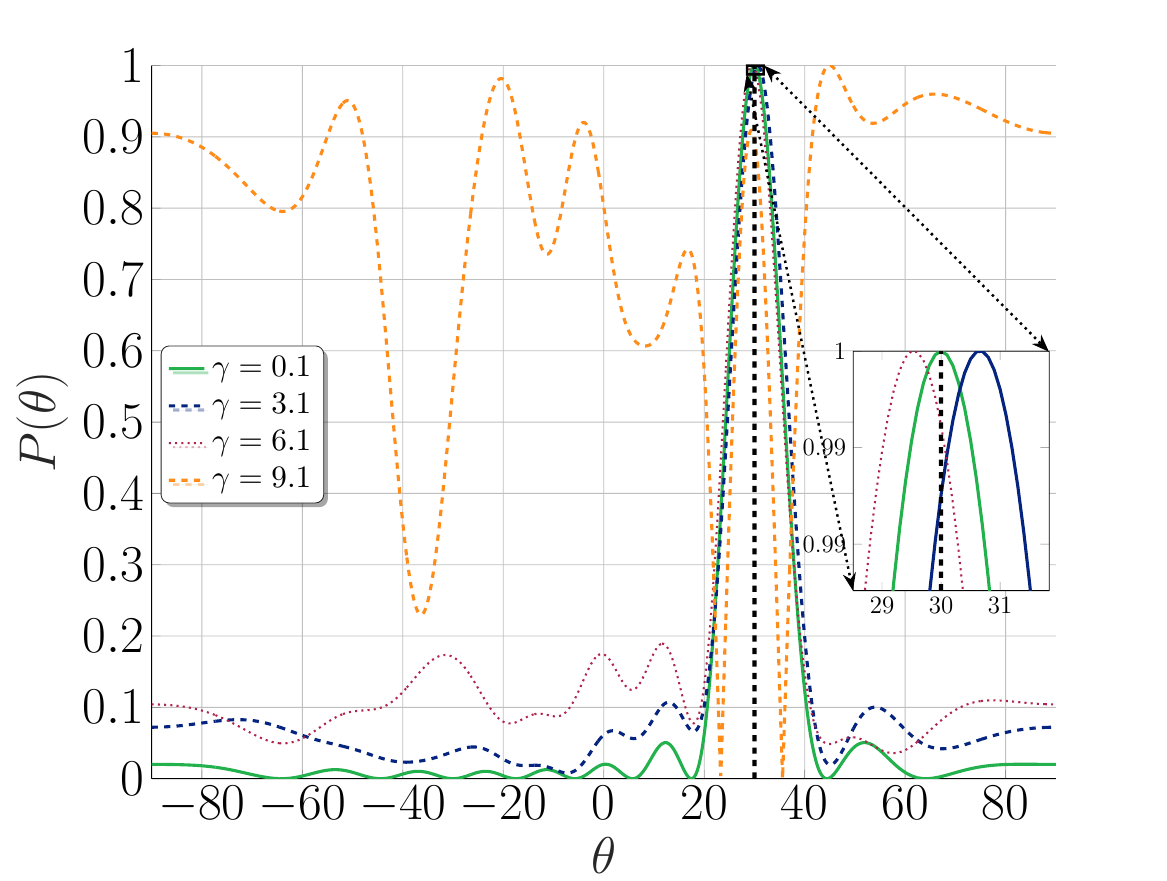} 
\label{fig:beampattern-01}}
\hfil
\subfloat[Target at $\theta_0 = 66^{\circ}$]{\includegraphics[height=2.25in,width=3.15in]{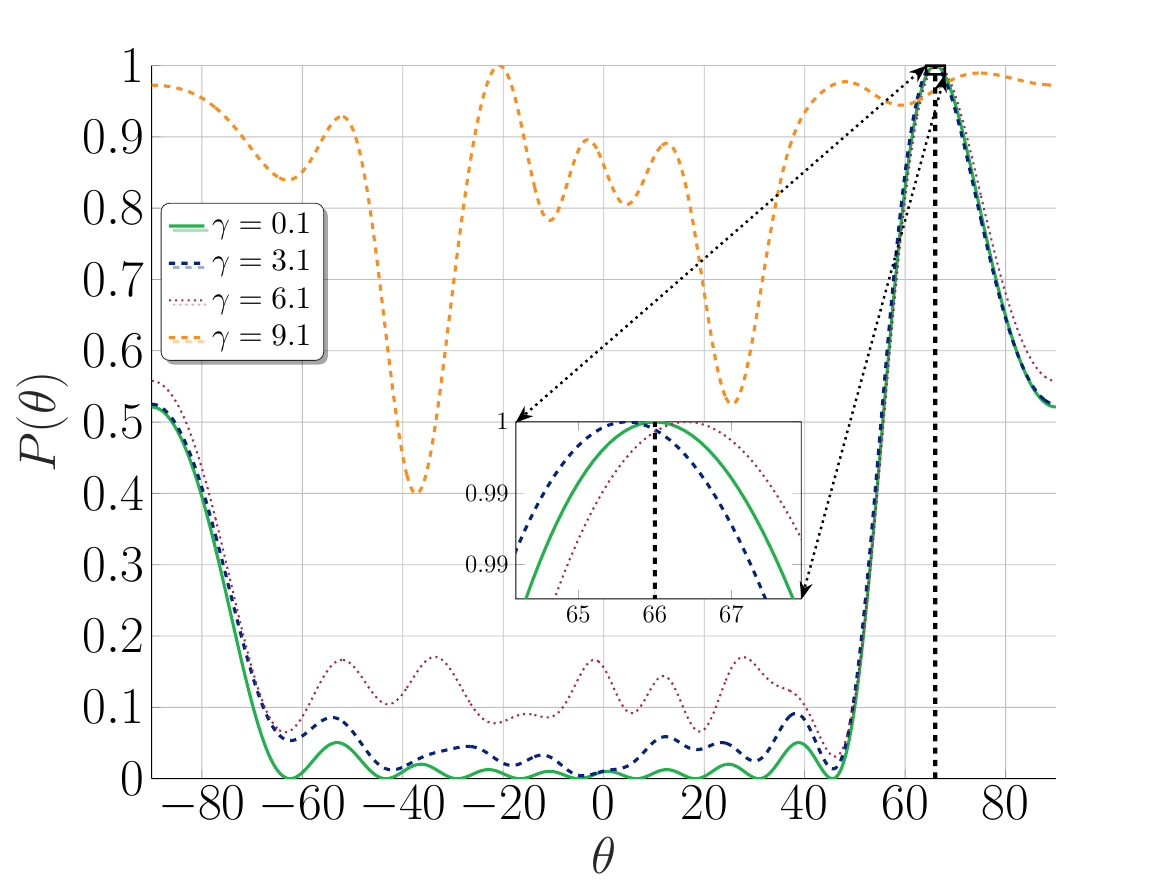}  
\label{fig:beampattern-02}}
\caption{Resulting radar beampatterns for different values of $\gamma$ at $N = 10, K = 4$ for different look directions, $\theta_0$.}
\label{fig:beampattern}
\end{figure*}

We also study several instances of the resulting beampattern through our design. It can be seen that not only lower sidelobes are achieved by lowering $\gamma$, but also a better accuracy in terms of look direction. For example, fixing $N = 10$ with $K = 4$ users, setting the look direction in $\theta_0 = 30^\circ$ as in Fig. \ref{fig:beampattern-01}, we can see that a decrementation of $\gamma$  by a factor of $3$ contributes to a sidelobe decrease of roughly $10 \dB$ and an improvement in the look direction in the order of $0.5^\circ$. A similar interpretation could be given in Fig. \ref{fig:beampattern-02}, where the look-direction has been changed to $\theta_0 = 66^\circ$.

\subsection{Validating SU BFing Expression}

\begin{figure}[!t]
\centering
\includegraphics[width=3.5in]{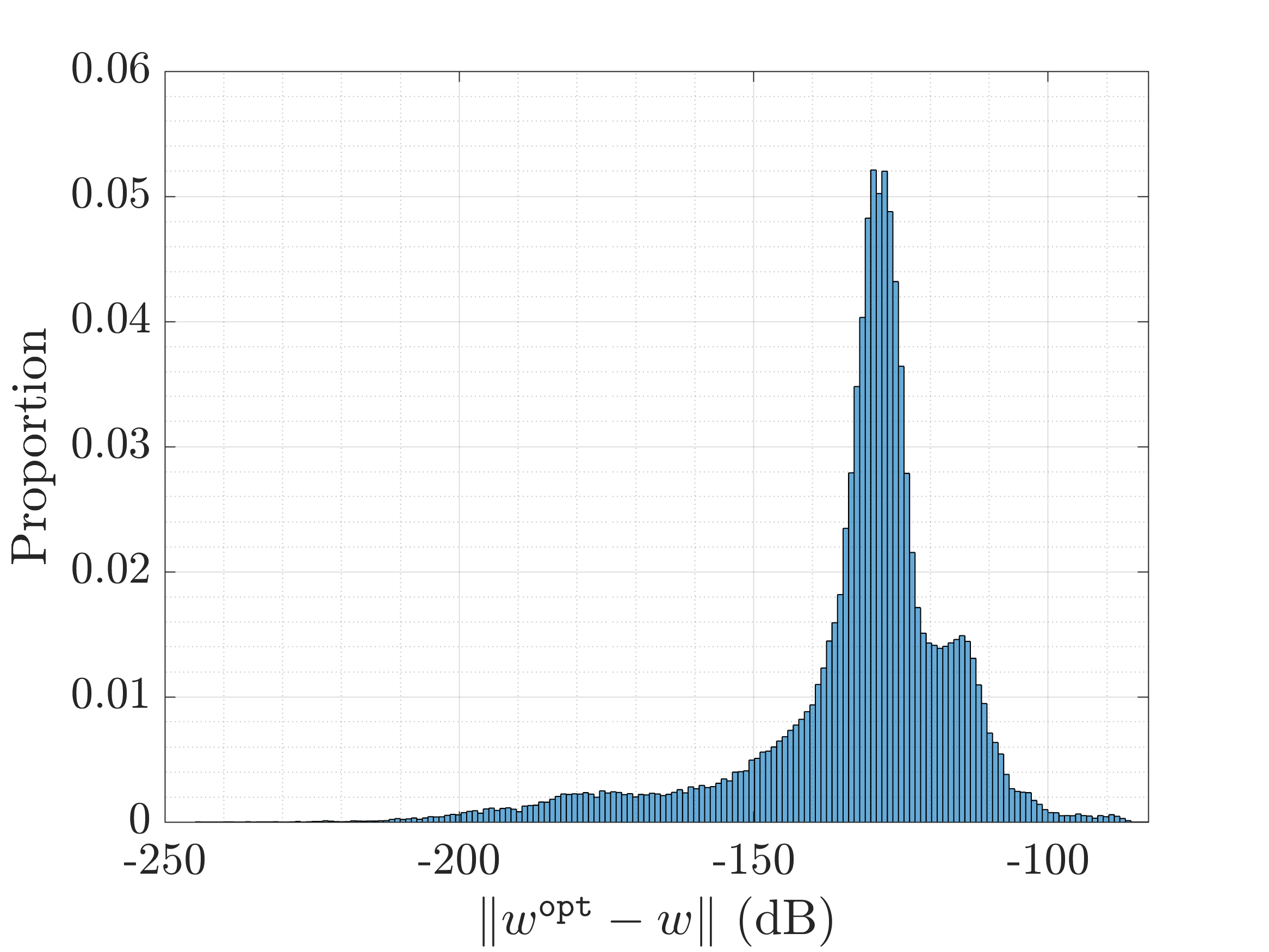}
\caption{Error distribution between the single user closed form solution, $\pmb{w}^{\opt}$ and the one given by $(\mathcal{P}_7)$ via CVX solver,  $\pmb{w}$.}
\label{fig:validating-su}
\end{figure}

In this subsection, we aim at validating the SU beamforming expressions obtained in \textbf{Theorem 3}. We have conducted $10^4$ Monte-Carlo trials by randomizing all parameters included within the optimization procedure, i.e. $\sigma_{\Delta},\pmb{h}$ and $\theta_0$. In Fig. \ref{fig:validating-su}, we depict a normalized histogram showing the distribution of errors (in dB) in the Frobenius norm sense, i.e. $\Vert \pmb{w}^{\opt} - \pmb{w} \Vert$, where $\pmb{w}^{\opt}$ is given in \textbf{Theorem 3} and $\pmb{w}$ is the solution of problem $(\mathcal{P}_7)$.  We see that the maximal error is bounded above by $-82$dB. The fluctuation of errors is due to fixed point numerical errors.

\subsection{Minimum Achievable Rate}
\begin{figure}[!t]
\centering
\includegraphics[width=3.5in]{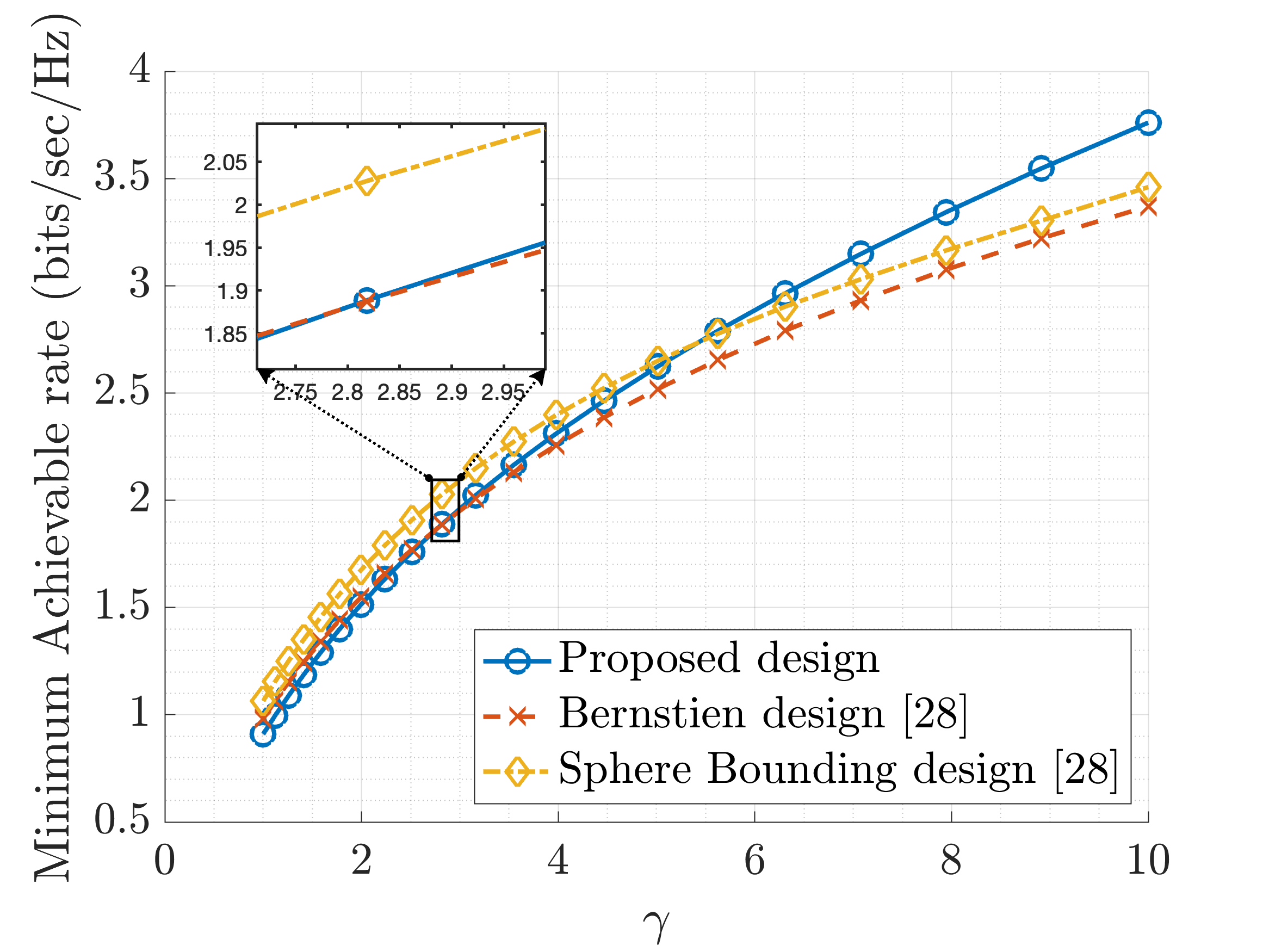}
\caption{The behaviour of the minimum achievable rate of the proposed method compared to the Bernstien and Sphere bounding designs \cite{GaussianRandomization}.}
\label{fig:MinachievableratePlusBenchmark}
\end{figure}
The minimum achievable rate is another meaningful metric due to the fact that the problem $(\mathcal{P})$ considers the QoS requirements per communication user. Futhermore, we also provide a comparison with state-of-the-art robust methods such as the Bernstein-type method and the sphere bounding approach for robust beamforming described in \cite{GaussianRandomization}. As shown in Fig. \ref{fig:MinachievableratePlusBenchmark}, we can observe that the proposed design and the Bernstein-type method coincide in terms of minimum achievable rate, when $\gamma < 2.8$. Interestingly, as $\gamma$ trespasses this value, the proposed design begins to outperform the Bernstien design method as $\gamma$ increases. Even more, when $\gamma > 5.6$ the proposed method also dominates the Sphere bounding method. It is worthy to note that both designs, i.e. the Bernstien and the Sphere Bounding methods aim at minimizing the total transmit beamforming power, given QoS SINR constraints. Therefore, both methods will target the minimum accepted rates with minimum power required. However, the proposed one targets the same rate with a given power budget and to maximize the output power of the Bartlett beamformer in a given direction.

\section{Conclusion}
\label{sec:conclusion}
In this paper, we have proposed an outage-based dual-functional radar-communication beamforming design aware of imperfect CSI, achieving high communication rates and detecting passive targets. To begin with, we have formulated an optimization problem that maximizes the output of the Bartlett beamformer towards the target, under outage signal-to-interference-plus-noise ratio constraints. Under some diligent relaxations, we arrive at a relaxed convex optimization problem, which can be tackled via numerical solvers, such as CVX. In terms of optimality, we prove that the final problem is always guaranteed to generate $\rank-$one beamforming solutions. Furthermore, we derive closed-form solutions of the problem for the single-user case, shed light on several interesting radar-communication trade-offs, and draw attention to gains, in the presence of imperfect CSI, achieved in radar through deliberate loss in communications, and vice-versa. Finally, and via numerical simulations, we show the superiority of the proposed beamforming design, as compared to state-of-the-art radar-communication beamformers.

Future research will be oriented towards joint sensing and communication beamforming and waveform design for multiple targets and imperfect CSI case. Other performance metrics for radar and communication could be taken into consideration. Nevertheless, future work will also consider estimation aspects of the radar subproblem and its impact on the communication performance.

\appendices
\small
\label{sec:appendix}
\section{Rank-1 Optimality}
\label{rankone}
We begin by writing down the Lagrangian function associated with problem $(\mathcal{P}_7)$, which is 
\begin{equation}
	\begin{split}
		&\mathcal{L}(\pmb{W}_1 \ldots \pmb{W}_K , \zeta , \pmb{a},\pmb{\alpha}, \pmb{\Psi},\pmb{Z})  \\
		&= \sum\limits_{k=1}^K
		\Tr(\Phi \pmb{W}_k ) +  \sum\limits_{k=1}^K
		\Tr (\pmb{Z}_k\pmb{Q}_k) + \sum\limits_{k=1}^K a_k \nu_k  \\
		&+ \sum\limits_{k=1}^K
		\alpha_k \Big( \Tr(\pmb{A}_k) - \sqrt{2 \epsilon_k}\mu_k -\epsilon_k \nu_k - \sigma_c^2 + {\pmb{h}}_k^T \bar{\pmb{W}}_{k} {\pmb{h}}_k^* \Big)\\
		&+  \sum\limits_{k=1}^K
		\Big(
		\nu_k \Tr(\pmb{\Psi}_k) +  \Tr(\pmb{\Psi}_k \pmb{A}_k)
		\Big)  
		-  \zeta  \Big(\sum\limits_{k=1}^K \Tr(\pmb{W}_k) - 1 \Big) ,
	 \end{split}
\end{equation}
where $\zeta  \geq 0$, $\alpha_k \geq 0 ,a_k \geq 0$ and $\pmb{\Psi}_k \succeq \pmb{0},\pmb{Z}_k \succeq \pmb{0}$ are Lagrangian dual variables. Let us define $\Phi = \pmb{a}(\theta_0)\pmb{a}^H(\theta_0)$ and let us partition $\pmb{Z}_k$ in the same way $\pmb{Q}_k$ is partitioned, namely, 
\begin{equation}
	\pmb{Z}_k
	=\begin{bmatrix}
		z_k  & 
	\begin{array}{cc}
		\pmb{\lambda}_k^H &
		\pmb{y}_k^H
	\end{array} \\
	\begin{array}{cc}
		\pmb{\lambda}_k \\
		\pmb{y}_k
	\end{array} &
	\pmb{\Pi}_k
	\end{bmatrix},
\end{equation} 
and so using trace properties of block matrices, we can write $\Tr(\pmb{Z}_k \pmb{Q}_k)$ as a quadratic function in  $ \sigma_{\Delta \pmb{h}_k}$ as follows
\begin{equation}
	\Tr(\pmb{Z}_k \pmb{Q}_k)
	= 
	\kappa_k + 
	\sqrt{2} \sigma_{\Delta \pmb{h}_k}
		\Tr(\bar{\pmb{W}}_k \pmb{\Upsilon}_k)+  \sigma_{\Delta \pmb{h}_k}^2 \Tr(\bar{\pmb{W}}_k \pmb{Y}_k),
\end{equation}
where $\kappa_k = \mu_kz_k + 
	\mu_k \Tr(\pmb{\Pi}_k)$ appears due to the diagonal contributions,  $\pmb{\Upsilon}_k = \pmb{h}_k^*\pmb{\lambda}_k^H + \pmb{\lambda}_k\pmb{h}_k^T$ and $\pmb{Y}_k =  \vect^{-1}(\pmb{y}_k)+\vect^{-H}(\pmb{y}_k)$ appear due to the first column/row of $\pmb{Q}_k,\pmb{Z}_k$, viz.
\begin{equation}
	\begin{split}
		\pmb{y}_k^H \vect(\pmb{A}_k) &+ \vect^H(\pmb{A}_k)\pmb{y}_k  \\
		&= \Tr(\vect^{-1}(\pmb{y}_k) \pmb{A}_k)+ \Tr(\vect^{-H}(\pmb{y}_k) \pmb{A}_k) \\
		&= \Tr(\pmb{A}_k \pmb{Y}_k) \\ &= \sigma_{\Delta \pmb{h}_k}^2 \Tr(\bar{\pmb{W}}_k \pmb{Y}_k),
	\end{split}	
\end{equation}
and
\begin{equation}
	\begin{split}
		\pmb{\lambda}_k^H \pmb{b}_k + \pmb{b}_k^H \pmb{\lambda}_k
		 &= 
		\sigma_{\Delta \pmb{h}_k}
		\big( \pmb{\lambda}_k^H \bar{\pmb{W}}_k \pmb{h}_k^*
		+
		\pmb{h}_k^T 
		\bar{\pmb{W}}_k \pmb{\lambda}_k \big) \\
		&=
		\sigma_{\Delta \pmb{h}_k}
		\Tr(\bar{\pmb{W}}_k \pmb{\Upsilon}_k).
	\end{split}
\end{equation}
Now the Lagrangian function could be expressed as follows
\begin{equation}
	\begin{split}
		& \mathcal{L}(\pmb{W}_1 \ldots \pmb{W}_K ,\zeta,  \pmb{a},\pmb{\alpha}, \pmb{\Psi},\pmb{Z}) \\ 
		&= \sum\limits_{k=1}^K
		\Tr(\Phi \pmb{W}_k ) +  
		\sum\limits_{k=1}^K 
		\Tr( \pmb{\Omega}_k \bar{\pmb{W}}_k) -
		\zeta \sum\limits_{k=1}^K \Tr(\pmb{W}_k) 
		\\ 
		&+ \sum\limits_{k=1}^K
		\alpha_k \Big(- \sqrt{2 \epsilon_k}\mu_k -\epsilon_k \nu_k - \sigma_c^2 \Big)\\
		&+  \sum\limits_{k=1}^K
		\Big(
		\nu_k \Tr(\pmb{\Psi}_k) + \kappa_k + a_k \nu_k
		\Big)  
		+
		\zeta,
	 \end{split}
\end{equation}
where $\pmb{\Omega}_k
	=
	\sqrt{2}  \sigma_{\Delta \pmb{h}_k} \pmb{\Upsilon}_k+ \sigma_{\Delta \pmb{h}_k}^2 (\pmb{Y}_k + \pmb{\Psi}_k) + \alpha_k ( \sigma_{\Delta \pmb{h}_k}^2\pmb{I}+{\pmb{h}}_k^* {\pmb{h}}_k^T )$. 
Using the expression of $\bar{\pmb{W}}_k = \frac{1}{\gamma_k}\pmb{W}_k - \sum\limits_{\ell \neq k} \pmb{W}_{\ell}$, we have
\begin{equation}
	\begin{split}
		&\mathcal{L}(\pmb{W}_1 \ldots \pmb{W}_K ,\zeta,  \pmb{a},\pmb{\alpha}, \pmb{\Psi},\pmb{Z})\\
		&= \sum\limits_{k=1}^K
		\Big(
		\Tr(\Phi \pmb{W}_k ) +  
		\frac{1}{\gamma_k}
	\Tr(\pmb{\Omega}_k \pmb{W}_k) -\Tr(\pmb{\Omega}_{\bar{k}}\pmb{W}_k) - \zeta  \Tr(\pmb{W}_k) \Big) \\ 
		&+ \sum\limits_{k=1}^K
		\alpha_k \Big(- \sqrt{2 \epsilon_k}\mu_k -\epsilon_k \nu_k - \sigma_c^2 \Big)\\
		&+  \sum\limits_{k=1}^K
		\Big(
		\nu_k \Tr(\pmb{\Psi}_k) + \kappa_k + a_k \nu_k
		\Big)   + \zeta ,
	 \end{split}
\end{equation}
where we have used $\sum\limits_{k=1}^K\sum\limits_{\ell \neq k}  \Tr(\pmb{\Omega}_k\pmb{W}_{\ell})  = \sum\limits_{k=1}^K\Tr(\pmb{\Omega}_{\bar{k}}\pmb{W}_k) $ where $\pmb{\Omega}_{\bar{k}} = \sum\limits_{\ell \neq k} \pmb{\Omega}_{\ell} $. The Lagrange dual function is then defined as follows
\begin{equation}
	g(\zeta, \pmb{a},\pmb{\alpha}, \pmb{\Psi},\pmb{Z}) = \max\limits_{ \lbrace \pmb{W}_k \succeq \pmb{0} \rbrace } \mathcal{L}(\pmb{W}_1 \ldots \pmb{W}_K , \zeta, \pmb{a},\pmb{\alpha}, \pmb{\Psi},\pmb{Z}).
\end{equation}
But since problem $(\mathcal{P}_7)$ is a convex optimization problem with zero duality gap, we can solve the problem via its dual, namely
\begin{equation}
(\mathcal{D}):
	\min\limits_{ \substack{\zeta \geq 0  \\ \lbrace a_k,\alpha_k \geq 0  \rbrace \\ 
							\lbrace \pmb{\Psi}_k,\pmb{Z}_k \succeq 0  \rbrace} } g(\zeta, \pmb{a},\pmb{\alpha}, \pmb{\Psi},\pmb{Z}).
\end{equation}
Let us denote the optimal solution of problem $(\mathcal{D})$ as $\zeta^{\opt},\pmb{a}^{\opt},\pmb{\alpha}^{\opt}, \pmb{\Psi}^{\opt},\pmb{Z}^{\opt}$ then the matrices $\pmb{W}_1^{\opt} \ldots \pmb{W}_K^{\opt} $ that maximize $\mathcal{L}(\pmb{W}_1 \ldots \pmb{W}_K ,\zeta^{\opt}, \pmb{a}^{\opt},\pmb{\alpha}^{\opt}, \pmb{\Psi}^{\opt},\pmb{Z}^{\opt})$ is the optimal solution of $(\mathcal{P}_7)$, which means we can find the optimal solution $\pmb{W}_1^{\opt} \ldots \pmb{W}_K^{\opt}$ thru the following problem
\begin{equation}
\label{FinalForm}
	 \max\limits_{  \pmb{W}_k \succeq \pmb{0}  }
	 \Tr(\Phi \pmb{W}_k ) -
	 \Tr \Bigg\lbrace  \bigg( \zeta^{\opt} \pmb{I} + \pmb{\Omega}_{\bar{k}}^{\opt} - \frac{1}{\gamma_k}\pmb{\Omega}_k^{\opt} \bigg) \pmb{W}_k \Bigg\rbrace,
\end{equation} 
where $\pmb{\Omega}_k^{\opt}
	=
	\sqrt{2}  \sigma_{\Delta \pmb{h}_k} \pmb{\Upsilon}_k^{\opt} + \sigma_{\Delta \pmb{h}_k}^2 (\pmb{Y}_k^{\opt} + \pmb{\Psi}_k^{\opt}) + \alpha_k ^{\opt}( \sigma_{\Delta \pmb{h}_k}^2\pmb{I}+{\pmb{h}}_k^* {\pmb{h}}_k^T )$, 
and  $\pmb{\Omega}_{\bar{k}}^{\opt} = \sum\limits_{\ell \neq k} \pmb{\Omega}_{\ell}^{\opt} $. Note that in equation \eqref{FinalForm}, all the constant terms have been omitted. In order for the problem in equation \eqref{FinalForm} to have a bounded value, it suffices to prove that matrix $\zeta^{\opt} \pmb{I}+  \pmb{\Omega}_{\bar{k}}^{\opt} - \frac{1}{\gamma_k}\pmb{\Omega}_k^{\opt}$ should be positive definite. Suppose that $ \zeta^{\opt} \pmb{I} + \pmb{\Omega}_{\bar{k}}^{\opt} - \frac{1}{\gamma_k}\pmb{\Omega}_k^{\opt}$ is not positive definite, then we can choose $\pmb{W}_k = p \pmb{w}\pmb{w}^H$ where $p > 0$ and so the problem could be written as 
\begin{equation}
\label{FinalForm-withContradiction}
	 \max\limits_{  \pmb{W}_k \succeq \pmb{0}  }
	 p \Big(  \vert  \pmb{a}^H(\theta_0) \pmb{w} \vert^2 -
	  \pmb{w}^H ( \zeta^{\opt} \pmb{I} + \pmb{\Omega}_{\bar{k}}^{\opt} - \frac{1}{\gamma_k}\pmb{\Omega}_k^{\opt} )  \pmb{w} \Big).
\end{equation} 
Now choose $\pmb{w}$ as a scaled multiple of one of the eigenvectors of $\zeta^{\opt} \pmb{I} + \pmb{\Omega}_{\bar{k}}^{\opt} - \frac{1}{\gamma_k}\pmb{\Omega}_k^{\opt}$ associated with its negative eigenvalues. This means that, as per the definition of eigenvalues/eigenvectors, along with the negative definiteness assumption, we have that $ \pmb{w}^H ( \zeta^{\opt} \pmb{I} + \pmb{\Omega}_{\bar{k}}^{\opt} - \frac{1}{\gamma_k}\pmb{\Omega}_k^{\opt}) \pmb{w} < 0 $, this means that $\vert \pmb{a}^H(\theta_0) \pmb{w} \vert^2 - \pmb{w}^H ( \zeta^{\opt} \pmb{I} + \pmb{\Omega}_{\bar{k}}^{\opt} - \frac{1}{\gamma_k}\pmb{\Omega}_k^{\opt} )  \pmb{w} > 0$ and so to maximize the entire cost appearing in equation \eqref{FinalForm-withContradiction}, it suffices to take $p \rightarrow +\infty$, therefore the problem becomes unbounded above, which is a contradiction of the optimality of $ \zeta^{\opt}, \pmb{a}^{\opt},\pmb{\alpha}^{\opt}, \pmb{\Psi}^{\opt},\pmb{Z}^{\opt}$. \\
Now that we have positive definiteness of $\zeta^{\opt} \pmb{I} +  \pmb{\Omega}_{\bar{k}}^{\opt} - \frac{1}{\gamma_k}\pmb{\Omega}_k^{\opt}$, then there is a unique Hermitian positive definite matrix $\pmb{R}_k^{\frac{1}{2}}$, such that $ \zeta^{\opt} \pmb{I}+ \pmb{\Omega}_{\bar{k}}^{\opt} - \frac{1}{\gamma_k}\pmb{\Omega}_k^{\opt} = \pmb{R}_k^{\frac{1}{2}}\pmb{R}_k^{\frac{1}{2}}$ \cite{Horn}. Using this square root matrix decomposition, we can express our cost in equation \eqref{FinalForm} as 
\begin{equation}
	 \max\limits_{  \widetilde{\pmb{W}}_k \succeq \pmb{0}  }
	  \big( \pmb{R}_k^{-\frac{1}{2}}\pmb{a}(\theta_0) \big)^H  \widetilde{\pmb{W}}_k  \big( \pmb{R}_k^{-\frac{1}{2}}\pmb{a}(\theta_0) \big)  -
	 \Tr \Big\lbrace \widetilde{\pmb{W}}_k \Big\rbrace ,
\end{equation} 
where $ \widetilde{\pmb{W}}_k = \pmb{R}_k^{\frac{1}{2}}\pmb{W}_k\pmb{R}_k^{\frac{1}{2}}$. Now, suppose that the optimal solution, denoted hereby as $\widetilde{\pmb{W}}_k^{\opt}$, is not $\rank$-one, i.e. $\rank$ $q > 1$. Then, employing the eigenvalue decomposition, we have that $\widetilde{\pmb{W}}_k^{\opt} = \sum\limits_{n=1}^q \beta_n \widetilde{\pmb{w}}_n \widetilde{\pmb{w}}_n^H$ where $\beta_n \geq 0$ and $\Vert \widetilde{\pmb{w}}_n \Vert = 1$, so the cost is upper bounded as follows
\begin{equation}
\begin{split}
&	\sum\limits_{n=1}^q \beta_n \Big\vert \big( \pmb{R}_k^{-\frac{1}{2}}\pmb{a}(\theta_0) \big)^H \widetilde{\pmb{w}}_n \Big\vert^2 
	- 
	\sum\limits_{n=1}^q \beta_n\\  \leq  &
	\sum\limits_{n=1}^q \beta_n \Big\vert \big( \pmb{R}_k^{-\frac{1}{2}}\pmb{a}(\theta_0) \big)^H \widetilde{\pmb{w}}_{\hat{n}} \Big\vert^2 
	- 
	\sum\limits_{n=1}^q \beta_n , 
\end{split}
\end{equation} 
where we have upper bounded all terms by $\Big\vert \big( \pmb{R}_k^{-\frac{1}{2}}\pmb{a}(\theta_0) \big)^H \widetilde{\pmb{w}}_{\hat{n}} \Big\vert$ where $\hat{n} = \argmax\limits_{n = 1 \ldots q} \Big\vert \big(  \pmb{R}_k^{-\frac{1}{2}}\pmb{a}(\theta_0) \big)^H \widetilde{\pmb{w}}_n \Big\vert $. But, this bound is achieved by the following $\rank$-one contribution $\widetilde{\pmb{W}}_k^{\rank-1} = \Bigg(\sum\limits_{n=1}^q \beta_n \Bigg) \widetilde{\pmb{w}}_{\hat{n}} \widetilde{\pmb{w}}_{\hat{n}}^H $, 
which is a contradiction of the assumption that $q > 1$. Therefore, $\widetilde{\pmb{W}}_k^{\opt} $ has to be $\rank$-1 optimal. Now, since we have ${\pmb{W}}_k^{\opt} = \pmb{R}_k^{-\frac{1}{2}}\widetilde{\pmb{W}}_k^{\opt}\pmb{R}_k^{-\frac{1}{2}}$, we must also have that ${\pmb{W}}_k^{\opt} $ is $\rank$-one optimal according to $\rank(\pmb{A}\pmb{B}) \leq \min\lbrace \rank(\pmb{A}),\rank(\pmb{B})\rbrace$ and the proof is done.
\section{Single User Closed-Form Solution}
\label{app:su-solution}
The following proof is trifold and goes as such:
\subsection{Part 1: Optimal solution spans $\lbrace \pmb{a}^*(\theta_0), \pmb{h}^* \rbrace$}
It is easy to see that when $\epsilon \rightarrow 0 $, the problem in \eqref{eq:problem1-linear-proba-bernstien-socp-SU2} converges to

\begin{equation}
\label{eq:problem1-linear-proba-bernstien-socp-SU3-epsilon-zero}
(\mathcal{P}_6^{\text{SU},\epsilon \rightarrow 0}):
\begin{aligned}
\begin{cases}
\max\limits_{ \pmb{w}}&   
\big\vert \pmb{a}^T(\theta_0) \pmb{w} \big\vert^2 \\
\textrm{s.t.}
 &  \gamma\sigma_c^2   - \sigma_{\Delta}^2 \Vert \pmb{w} \Vert^2  \leq    \vert {\pmb{h}}^T \pmb{w}  \vert^2 \\
     & \Vert \pmb{w} \Vert^2 \leq 1 \\ 
\end{cases}
\end{aligned}
\end{equation}
First, we prove that the optimal solution of the above problem has to be a linear combination of $\pmb{a}^*(\theta_0)$ and $\pmb{h}^*$. In other words, $\pmb{w}^{\opt} \in \Span(\pmb{a}^*(\theta_0),\pmb{h}^*)$. To prove this, let us express any feasible $\pmb{w}$ as a decomposition of two vectors, namely $\pmb{w} = \varphi \pmb{w}_{\varphi} + \varrho \pmb{w}_{\varrho} $, 
where $\Vert \pmb{w}_{\varphi} \Vert = \Vert \pmb{w}_{\varrho} \Vert = 1$, $\pmb{w}_{\varphi} \perp \pmb{w}_{\varrho}$ and $\pmb{w}_{\varphi} \in  \Span(\pmb{a}^*(\theta_0),\pmb{h}^*)$, i.e. $\pmb{w}_{\varrho}$ lies in the null space of $\Span(\pmb{a}^*(\theta_0),\pmb{h}^*)$. Note that $\varrho$ does not contribute to an increase in the cost because $\pmb{a}^T(\theta_0) \pmb{w}_{\varrho} = 0$. Therefore, cost maximization is attained by increasing $\varphi$, only. Next, a feasible solution in terms of $\varrho,\varphi$ should satisfy
\begin{equation}
\label{firstconstraint01}
	\gamma\sigma_c^2   - \sigma_{\Delta}^2 ( \varphi^2 +  \varrho^2)  \leq    \varphi^2 \vert {\pmb{h}}^T \pmb{w}_{\varphi}  \vert^2,
\end{equation}
and $\varphi^2 +  \varrho^2 \leq 1$, 
where the upper bound in \eqref{firstconstraint01} appears due to $\pmb{h}^T \pmb{w}_{\varrho} = 0$. For a given set of parameters, $(\gamma,\sigma_c^2,\sigma_{\Delta}^2)$, it can be easily seen that an increase in $\varphi$ would only increase the gap between the two bounds in the constraint appearing in equation \eqref{firstconstraint01}. Therefore, to sum things up, an increase of $\varphi$ only would increase the cost, while preserving feasibility of the solution. Therefore, to achieve the most out of the problem, one could choose $\varphi = 1$ and $\varrho = 0$, which means that $\pmb{w}$ lives in the subspace spanned by $\pmb{a}^*(\theta_0)$ and $\pmb{h}^*$.\\
\subsection{Part 2: Optimal solution expressed in $\lbrace \pmb{a}_{\parallel}, \pmb{a}_{\perp} \rbrace$}
\label{app:subspace-a}
Now that we proved that the optimal solution is within $\Span(\pmb{a}^*(\theta_0),\pmb{h}^*)$, then let's consider an orthonormal representation of that subspace, which could be done via the Gram-Schmidt process, i.e. let's pick the axis of reference first to be $\pmb{a}^*(\theta_0)$, i.e. $\pmb{a}_{\parallel} = \frac{\pmb{a}^*(\theta_0)}{\Vert \pmb{a}(\theta_0)\Vert}$ and $\pmb{a}_{\perp} = \frac{\pmb{h}^* - \pmb{a}_{\parallel}^H\pmb{h}^*\pmb{a}_{\parallel}  }{\Vert\pmb{h}^* - \pmb{a}_{\parallel}^H\pmb{h}^*\pmb{a}_{\parallel}  \Vert}$.
Let the optimal solution in this case be written as $\pmb{w}^{\opt}
	=
	\alpha_{\parallel} \pmb{a}_{\parallel}
	+ 
	\alpha_{\perp} \pmb{a}_{\perp}$.
Replacing this representation in the optimization problem of equation \eqref{eq:problem1-linear-proba-bernstien-socp-SU3-epsilon-zero}, we get 
\begin{equation}
\label{eq:problem1-linear-proba-bernstien-socp-GramSchmdt1}
(\mathcal{P}_6^{\text{SU},\epsilon \rightarrow 0}):
\begin{aligned}
\begin{cases}
\max\limits_{ \alpha_{\parallel} , \alpha_{\perp} }&   
\vert \alpha_{\parallel} \vert^2 \overbrace{\big\vert  \pmb{a}^T(\theta_0)\pmb{a}_{\parallel}\big\vert^2}^N \\
\textrm{s.t.}
 &  \gamma\sigma_c^2   - \sigma_{\Delta}^2(\vert \alpha_{\parallel} \vert^2 + \vert \alpha_{\perp} \vert^2 ) \leq    
 \vert {\pmb{h}}^T \pmb{w}^{\opt} \vert^2 \\
     & \vert \alpha_{\parallel} \vert^2 + \vert \alpha_{\perp} \vert^2 \leq 1 \\ 
\end{cases}
\end{aligned}
\end{equation}
It is clear that from the above, the cost function increases as a function of $\vert \alpha_{\parallel} \vert^2 $, with no contribution from $\alpha_{\perp}$. To satisfy the unit circle constraint, take $\vert \alpha_{\parallel} \vert^2 = 1 $ and $\vert \alpha_{\perp} \vert^2 = 0$. Now, so that this solution, namely $\pmb{w}^{\opt} = \pmb{a}_{\parallel}$ is optimal, the first constraint in problem \eqref{eq:problem1-linear-proba-bernstien-socp-GramSchmdt1} needs to be feasible, that is 
\begin{equation}
\label{eq:the-epic-condition}
	\gamma\sigma_c^2   - \sigma_{\Delta}^2 \leq     
 \vert {\pmb{h}}^T \pmb{a}_{\parallel} \vert^2 .
\end{equation}
Using the representation of $\pmb{a}_{\parallel}$, the condition in equation \eqref{eq:the-epic-condition} becomes
\begin{equation}
\label{eq:the-epic-condition-2}
	N(\gamma\sigma_c^2   - \sigma_{\Delta}^2 ) \leq     
 \vert {\pmb{h}}^T \pmb{a}^*(\theta_0) \vert^2 .
\end{equation}
Also note that by Cauchy-Schwarz, we have that $N(\gamma\sigma_c^2   - \sigma_{\Delta}^2 ) \leq \Vert \pmb{h} \Vert^2  N$, hence the factor $\frac{\gamma\sigma_c^2   - \sigma_{\Delta}^2}{\Vert \pmb{h} \Vert^2} \leq 1$, which gives us a bound on the choice of parameters $(\gamma,\sigma_{\Delta},\sigma_c^2)$ as $\gamma \leq \frac{\Vert \pmb{h} \Vert^2 + \sigma_{\Delta}^2}{\sigma_c^2}$, which also happens to be the upper bound on single user average-$\SINR$, under a unit power budget constraint on $\pmb{w}$. Indeed
\begin{equation}
\label{eq:max-achieve}
\begin{split}
	\max_{\Vert \pmb{w} \Vert  = 1}
	\mathbb{E} \lbrace \SINR \rbrace 
	&=
	\max_{\Vert \pmb{w} \Vert  = 1}
	\frac{ \pmb{w}^H  \mathbb{E} \lbrace \tilde{\pmb{h}}^*\tilde{\pmb{h}}^T \rbrace \pmb{w}   } 
	     {\sigma_{c}^2} \\
	&=
	\max_{\Vert \pmb{w} \Vert  = 1}
	\frac{ \pmb{w}^H  ({\pmb{h}}^*{\pmb{h}}^T + \sigma_{\Delta}^2 \pmb{I} ) \pmb{w}   } 
	     {\sigma_{c}^2} \\
	&= 
	\frac{\Vert \pmb{h} \Vert^2 + \sigma_{\Delta}^2}{\sigma_c^2},
\end{split}
\end{equation}
where the last step is by choosing $\pmb{w}$ to be the eigenvector corresponding to the maximum eigenvalue of ${\pmb{h}}^*{\pmb{h}}^T + \sigma_{\Delta}^2 \pmb{I}$.

\subsection{Part 3: Optimal solution expressed in $\lbrace \pmb{h}_{\parallel}, \pmb{h}_{\perp} \rbrace$}
Using the same approach as in Appendix \ref{app:subspace-a}, we consider an orthonormal representation of the subspace $\Span( \pmb{a}^*(\theta_0), \pmb{h}^*)$, but instead we now pick the reference to be $\pmb{h}^*$, viz. $\pmb{h}_{\parallel} = \frac{\pmb{h}^*}{\Vert \pmb{h} \Vert}$ and $\pmb{h}_{\perp} = \frac{\pmb{a}^*(\theta_0) - \pmb{h}_{\parallel}^H\pmb{a}^*(\theta_0)\pmb{h}_{\parallel}  }{\Vert \pmb{a}^*(\theta_0) - \pmb{h}_{\parallel}^H\pmb{a}^*(\theta_0)\pmb{h}_{\parallel}\Vert}$, 
and so the solution takes the form $\pmb{w}^{\opt}
	=
	\alpha_1 \pmb{h}_{\parallel}
	+ 
	\alpha_2 \pmb{h}_{\perp}$.
Replacing this representation in the optimization problem of equation \eqref{eq:problem1-linear-proba-bernstien-socp-SU3}, we get 

\begin{equation}
\label{eq:problem1-linear-proba-bernstien-socp-SU3}
(\mathcal{P}_6^{\text{SU},\epsilon \rightarrow 0}):
\begin{aligned}
\begin{cases}
\max\limits_{ \alpha_1 , \alpha_2 }&   
\big\vert \alpha_1 \pmb{a}^T(\theta_0) \pmb{h}_{\parallel} + \alpha_2 \pmb{a}^T(\theta_0) \pmb{h}_{\perp} \big\vert^2 \\
\textrm{s.t.}
 &  \gamma\sigma_c^2   - \sigma_{\Delta}^2(\vert \alpha_1 \vert^2 + \vert \alpha_2 \vert^2 ) \leq    
  \vert \alpha_1 \vert^2\Vert {\pmb{h}}   \Vert^2 \\
     & \vert \alpha_1 \vert^2 + \vert \alpha_2 \vert^2 \leq 1. \\ 
\end{cases}
\end{aligned}
\end{equation}

It is clear that the cost is maximized upon adjusting the phases of $\alpha_1,\alpha_2$ as follows

\begin{align}
	\exp(j\angle \alpha_{1,2} )&= \exp \lbrace - j \angle (\pmb{a}^T(\theta_0) \pmb{h}_{\parallel,\perp}) \rbrace = \frac{\pmb{h}_{\parallel,\perp}^H \pmb{a}^*(\theta_0)}{\Vert \pmb{h}_{\parallel,\perp}^H \pmb{a}^*(\theta_0) \Vert}.
\end{align}
In that case, both complex numbers $ \alpha_1 \pmb{a}^T(\theta_0) \pmb{h}_{\parallel}$ and $ \alpha_2 \pmb{a}^T(\theta_0) \pmb{h}_{\perp}$ add up constructively. Then, it is clear that the gains $\vert \alpha_1 \vert$ and $\vert \alpha_2 \vert$ should be jointly maximized, therefore, the last constraint should be attained with equality, i.e. $ \vert \alpha_1 \vert^2 + \vert \alpha_2 \vert^2 = 1$. In that case, the first constraint translates to $ 1 \geq \vert	\alpha_1 \vert^2 \geq \rho $ and $ \vert	\alpha_2 \vert^2 \leq 1- \rho \leq 1$, where $0 \leq \rho \eqdef \frac{\gamma \sigma_c^2 - \sigma_{\Delta}^2 }{\Vert \pmb{h} \Vert^2} \leq 1$, 
which is satisfied upon $\gamma \leq \frac{\Vert \pmb{h} \Vert^2 + \sigma_{\Delta}^2}{\sigma_c^2}$. Once $\vert \alpha_1 \vert$ and $\vert \alpha_2 \vert$ are set, then we have an optimal solution. Working in the compliment region of that defined in equation \eqref{eq:the-epic-condition-2}, i.e. $N(\gamma\sigma_c^2   - \sigma_{\Delta}^2 ) >    
 \vert {\pmb{h}}^T \pmb{a}^*(\theta_0) \vert^2$, 
or equivalently $\vert \pmb{a}^T(\theta_0) \pmb{h}_{\parallel} \vert^2 \leq N \rho$. Also since $\pmb{h}_{\parallel},\pmb{h}_{\perp} $ form an orthonormal basis in $\Span (\pmb{a}^*(\theta_0), \pmb{h}^*)$, which means $\vert \pmb{a}^T(\theta_0) \pmb{h}_{\parallel} \vert^2 + \vert \pmb{a}^T(\theta_0) \pmb{h}_{\perp}\vert^2 = \Vert \pmb{a}^T(\theta_0) \Vert^2 = N $. Based on this argument, one also has $\vert \pmb{a}^T(\theta_0) \pmb{h}_{\perp}\vert^2 \geq N (1-\rho)$. It could be shown that by the choice of $\vert \alpha_1 \vert^2 = \rho$ and so $\vert \alpha_2 \vert^2 = 1- \rho$ maximizes the cost. Finally, the optimal solution is
\begin{equation}
	\pmb{w}^{\opt}
	=
	\underbrace{\sqrt{\rho}\frac{\pmb{h}_{\parallel}^H \pmb{a}^*(\theta_0)}{\Vert \pmb{h}_{\parallel}^H \pmb{a}^*(\theta_0) \Vert}}_{\alpha_1} \pmb{h}_{\parallel} + 
	\underbrace{\sqrt{1-\rho}\frac{\pmb{h}_{\perp}^H \pmb{a}^*(\theta_0)}{\Vert \pmb{h}_{\perp}^H \pmb{a}^*(\theta_0) \Vert}}_{\alpha_2}\pmb{h}_{\perp} .
\end{equation}

\section*{Acknowledgment}
The authors would like to thank the anonymous reviewers for their constructive comments, which contributed in improving the manuscript. The authors would also like to thank Dr. Lasha Ephremidze and Prof. Ilya Spitkovsky of the NYU Abu Dhabi Mathematics department for help with the proofs of the theorems appearing in this manuscript.

\end{document}